\documentclass[aps,prd,floatfix,nofootinbib,showpacs,
12pt]{revtex4-1}

\usepackage{amssymb}
\usepackage[intlimits]{amsmath}
\usepackage{amsfonts}
\usepackage{dsfont}
\usepackage{subfigure}
\usepackage[usenames,dvipsnames]{color}
\usepackage{graphicx}
\usepackage{natbib}
\usepackage{multirow}
\usepackage{bbm}

\usepackage{slashed}
\usepackage{nicefrac}

\newcommand{\Dslash}{\ensuremath{D\hspace{-1.5ex} /}}
\newcommand{\DslashIndex}{\ensuremath{D\hspace{-1ex} /}}
\newcommand{\Tr}{\ensuremath{\operatorname{Tr}}}

\newcommand{\Det}{\ensuremath{\operatorname{Det}}}

\newcommand{\be}{\begin{equation}}
\newcommand{\ee}{\end{equation}}
\newcommand{\bea}{\begin{eqnarray}}
\newcommand{\eea}{\end{eqnarray}}
\newcommand{\bi}{\begin{itemize}}
\newcommand{\ei}{\end{itemize}}

\newcommand{\xmark}{{\text{\sffamily X}}}

\newcommand{\gb}{\bar{g}}
\newcommand{\Db}{\bar{D}}
\newcommand{\Rb}{\bar{R}}

\newcommand{\cJ}{\mathcal{J}}
\newcommand{\cO}{\mathcal{O}}
\newcommand{\cR}{\mathcal{R}}
\newcommand{\cV}{\mathcal{V}}
\newcommand{\cT}{\mathcal{T}}

\begin{document}
\title{Asymptotically safe $f(R)$-gravity coupled to matter I: \\
the polynomial case}

\author {Nat\'alia Alkofer}
\email{N.Alkofer@science.ru.nl}
\author {Frank Saueressig}
\email{F.Saueressig@science.ru.nl}

\affiliation{ Institute for Mathematics, 
Astrophysics, and Particle Physics (IMAPP), Radboud University Nijmegen, \\
Heyendaalseweg 135, 6525 AJ Nijmegen, The Netherlands}

\date{\today}


\begin{abstract}
        We use the functional renormalization group equation for the effective average action to study the non-Gaussian renormalization group fixed points (NGFPs) arising within the framework of $f(R)$-gravity minimally coupled to an arbitrary number of scalar, Dirac, and vector fields. Based on this setting we provide comprehensible estimates which gravity-matter systems give rise to NGFPs suitable for rendering the theory asymptotically safe. The analysis employs an exponential split of the metric fluctuations and retains a 7-parameter family of coarse-graining operators allowing the inclusion of non-trivial endomorphisms in the regularization procedure. For vanishing endomorphisms, it is established that gravity coupled to the matter content of the standard model of particle physics and many beyond the standard model extensions exhibit NGFPs whose properties are strikingly similar to the case of pure gravity: there are two UV-relevant directions and the position and critical exponents converge rapidly when higher powers of the scalar curvature are included. Conversely, none of the phenomenologically interesting gravity-matter systems exhibits a stable NGFP when a Type II coarse graining operator is employed. Our analysis resolves this tension by demonstrating that the NGFPs seen in the two settings belong to different universality classes.
\end{abstract}

\maketitle
\section{Introduction}
\label{sec:intro}
A unified description of all fundamental forces, valid on all length scales, is a major open problem in theoretical physics. Quite surprisingly, the asymptotic safety mechanism, first suggested by Weinberg in the context of gravity \cite{Weinberg,Weinberg:1996kw}, may lead to such a description within the well-established framework of quantum field theory. The key ingredients underlying this idea are non-Gaussian (or interacting) fixed points (NGFPs) of the {{theories' renormalization group (RG) flow.} A theory whose high-energy behavior is controlled by such a fixed point is free from unphysical ultraviolet (UV) divergences. Moreover, the predictive power provided by the NGFP may be comparable to the one known from perturbatively renormalizable quantum field theories, see \cite{Niedermaier:2006wt,Codello:2008vh,Litim:2011cp,Percacci:2011fr,Reuter:2012id,Nagy:2012ef,Percacci:2017fkn} for reviews. 

Starting from the seminal work \cite{Reuter:1996cp} the existence of a suitable NGFP for gravity has been demonstrated in increasingly sophisticated approximations starting from the Einstein-Hilbert action \cite{Souma:1999at,Falkenberg:1996bq,Reuter:2001ag,Lauscher:2001ya,Litim:2003vp,
  Bonanno:2004sy,Eichhorn:2009ah,Manrique:2009uh,
  Eichhorn:2010tb,Groh:2010ta,Manrique:2010am,Christiansen:2012rx,Codello:2013fpa,
  Christiansen:2014raa,Becker:2014qya,Falls:2014zba,
  Falls:2015qga,Christiansen:2015rva,
  Gies:2015tca,Benedetti:2015zsw,
  Pagani:2016dof,Denz:2016qks,Falls:2017cze,
  Knorr:2017fus}, 
its extension by higher-derivative and higher-order curvature terms \cite{Lauscher:2002sq,Reuter:2002kd,
  Codello:2006in,Codello:2007bd,Machado:2007ea,
  Niedermaier:2009zz,Benedetti:2009rx,Benedetti:2009gn,Benedetti:2009iq,Benedetti:2010nr,Rechenberger:2012pm,
  Ohta:2013uca,Falls:2013bv,Benedetti:2013jk,
  Falls:2014tra,
  Eichhorn:2015bna,Ohta:2015efa,
  Falls:2016wsa,Falls:2016msz,Christiansen:2016sjn,
  Gonzalez-Martin:2017gza,Becker:2017tcx}, the construction of fixed functions including an infinite number of coupling constants \cite{Reuter:2008qx,Benedetti:2012dx,Demmel:2012ub,Dietz:2012ic,Bridle:2013sra,Dietz:2013sba,Demmel:2014sga,Demmel:2014hla,Demmel:2015oqa,Ohta:2015fcu,Labus:2016lkh,Dietz:2016gzg,Christiansen:2017bsy,Knorr:2017mhu,Falls:2017lst}, up to including the notorious Goroff-Sagnotti two-loop counterterm \cite{Gies:2016con}. This NGFP also persists in the presence of a foliation structure \cite{Manrique:2011jc,Rechenberger:2012dt,Biemans:2016rvp,Biemans:2017zca,Houthoff:2017oam}.

Based on the successes for pure gravity, it is natural to also incorporate matter degrees of freedom. While it is commonly established that the realization of the asymptotic safety mechanism within gravity-matter systems leads to bounds on the admissible number of matter fields and their interactions 
\cite{Dou:1997fg,Percacci:2002ie,Narain:2009fy,Daum:2010bc,
  Folkerts:2011jz,Harst:2011zx,Eichhorn:2011pc,
  Eichhorn:2012va,Dona:2012am,Henz:2013oxa,Dona:2013qba,
  Percacci:2015wwa,Labus:2015ska,Oda:2015sma,Meibohm:2015twa,Dona:2015tnf,
  Meibohm:2016mkp,Eichhorn:2016esv,Henz:2016aoh,Eichhorn:2016vvy,
  Christiansen:2017gtg,Eichhorn:2017eht,Christiansen:2017qca,
  Eichhorn:2017ylw,Eichhorn:2017lry,Eichhorn:2017egq,Christiansen:2017cxa}, the complete picture is far from clear. In particular, a systematic study of the predictive power of the gravity-matter fixed points along the lines of $f(R)$-gravity, which played a pivotal role in the case of pure gravity, is still missing. The goal of the present work is to provide this analysis.\footnote{For some related work in this direction also see \cite{Schroder:2015xva}.}

We obtain the flow equation for $f(R)$-gravity minimally coupled to an arbitrary number of minimally coupled matter fields by a suitable projection of Wetterich's equation \cite{Wetterich:1992yh,Morris:1993qb} adapted to the case of gravity \cite{Reuter:1996cp}
\be\label{FRGE}
\partial_t \Gamma_k = \frac 1 2 \, {\mathrm {STr}} \left[ \left(\Gamma_k^{(2)} + 	\cR_k\right)^{-1} \partial_t \cR_k \right] \, .  
\ee
Here $t = \ln(k/k_0)$ is the RG time with $k_0$ being an arbitrary reference scale, $\Gamma_k$ denotes the effective average action, $\Gamma_k^{(2)}$ its second variation with respect to the fluctuation fields and $\cR_k$ is a mass-type regulator which ensures that the flow is governed by integrating out quantum fluctuations with eigenvalues $\lambda \approx k^2$. In order to facilitate the comparison with previous studies our technical implementation closely follows \cite{Ohta:2015efa,Ohta:2015fcu}. In particular we work on a $d$-dimensional background sphere so that the operator-valued traces can be computed as sums over eigenvalues of the corresponding differential operators. In this way the construction incorporates a $7$-parameter family of coarse graining operators. The free (endomorphism) parameters implement relative shifts in the eigenvalue spectra. Essentially, they determine which fluctuating modes are integrated out to drive the flow at the scale $k^2$. The consequences of using different coarse graining operators can then be studied systematically.

As a key result, our work establishes that $f(R)$-gravity coupled to matter gives rise to \emph{two different families of universality classes}. While these classes are virtually indistinguishable at the level of the Einstein-Hilbert truncation, the inclusion of higher-derivative operators (foremost the $R^2$-terms) reliably disentangles the two families. The ``gravity''-type family has a stable extension under the inclusion of higher-derivative operators and the expansion exhibits a rapid convergence in terms of the position of the fixed point and its critical exponents. Moreover, the family comes with a low number of free parameters associated with relevant operators. Provided that the coarse graining operator is chosen as the Laplacian, it is found that many phenomenologically interesting gravity-matter systems actually possess a NGFP belonging to this family, see Table \ref{Tab.mainresults}. In contrast, the ``matter-dominated'' family of fixed points turns out to be unstable under the inclusion of higher-derivative terms. It is then shown explicitly, that changing the coarse graining operator may have the effect of leaving the domain supporting the ``gravity-type'' fixed points replacing them by a fixed point from the matter-dominated family, see Fig.\ \ref{Fig.typeIIannihilation}. While this does not provide insights on how the coarse graining operator should be chosen, it nevertheless offers a natural explanation for the qualitatively different results on gravity-matter systems in the literature: depending on the precise setup, the computation may probe a fixed point belonging to either of the two distinguished families. If this fixed point happens to be of ``matter-dominated'' type one may then expect that the fixed point may become unstable once the approximation exceeds a certain degree of sophistication. We expect that this feature also persists in computations which resolve background and fluctuation vertices of the effective average action.
 
 The rest of the work is organized as follows. Sect.\ \ref{sec:Gamma} introduces our general setting and constructs the explicit operator traces appearing in the flow equation \eqref{FRGE}. Relying on the average interpolation for the spectral sums used in \cite{Ohta:2015efa,Ohta:2015fcu}, Sect.\ \ref{sec:spectrum} describes the evaluation of these traces leading to the partial differential equation \eqref{pdf4d} governing the scale-dependence of the function $f(R)$. The fixed point structure encoded in this flow equation is investigated in Sect.\ \ref{sec:fixedpoints} and we briefly summarize our findings in Sect.\ \ref{sec:conclusions}. Appendix \ref{App.A} and \ref{App.B} provide technical details about the construction of spectral sums and a detailed survey of the fixed points found for phenomenologically interesting gravity-matter systems, respectively.

\section{Flow equation for gravity-matter systems in the $f(R)$-truncation}
\label{sec:Gamma}
In this section we introduce our ansatz for the effective average action (Sect.\ \ref{sec:setup}) and construct the operator traces entering into our projected flow equation (Sect.\ \ref{sec:RGgrav}). The properties of the regularization schemes employed in this work are discussed in Sect.\ \ref{sec:endomorphism}.
\subsection{Setup}
\label{sec:setup}
In this work we study the RG flow of $f(R)$-gravity supplemented by minimally coupled matter fields. 
Throughout the work we resort to a  $\Gamma_k$ of the form
\be\label{Gammaans}
\Gamma_k = \Gamma_k^{\rm grav} + \Gamma^{\rm matter}_k \, ,
\ee
where $\Gamma^{\rm grav}_k$ is the gravitational part of the effective average action and $\Gamma^{\rm matter}$ encodes the contribution of the matter fields. The gravitational part of the action is given by
\be
\Gamma_k^{\rm grav} = \int d^dx \, \sqrt{g} \, f_k(R)  \,\, + \Gamma_k^{\rm gf} + \Gamma_k^{\rm gh}  \,  ,
\label{GammaGrav} 
\ee
where $f_k(R)$ is an arbitrary, scale-dependent function of the Ricci scalar $R$ and the action
is supplemented by suitable gauge fixing and ghost terms. This sector is taken to be identical to the one
studied in \cite{Ohta:2015fcu}. The matter sector, $
 \Gamma^{\rm matter}  =  \Gamma^{\rm scalar}  +  \Gamma^{\rm fermion}  +  \Gamma^{\rm vector}$, 
contains $N_S$ scalar fields $\phi$, $N_D$ Dirac fermions $\psi$, and $N_V$ abelian gauge fields $A_\mu$ (including appropriate gauge fixing to Feynman gauge and ghosts $\bar{c},c$). Their actions are given by
\begin{subequations}  \label{mattersector}
	\begin{align} \label{Gscalar}
	\Gamma^{\rm scalar}  = &  
	\frac{N_S}{2} \int d^d x \sqrt{g} g^{\mu\nu} \left(  D_\mu \phi \right) \left( D_\nu \phi \right) \, ,  \\ \label{Gfermion}
	\Gamma^{\rm fermion} = &  \, i \, N_D \,  \int d^d x \sqrt{g} \, \bar \psi  \Dslash \, \psi \, , \\ \label{Gvector}
\Gamma^{\rm vector} = & N_V  \int d^d x \sqrt{g} \, 
\left( \frac 1 4 F^{\mu\nu} F_{\mu \nu}
 + \frac 1 2 ( D^\mu A_\mu)^2 
+ \bar c \, (- D^2) \, c
\right) \, .  
	\end{align}
\end{subequations}

The construction of the flow equation \eqref{FRGE} employs the background field method. Following \cite{Ohta:2015fcu}, we implement an exponential split of the physical metric $g_{\mu\nu}$ into a fixed background $\gb_{\mu\nu}$ and fluctuations $h_{\mu\nu}$
\be\label{ExpSplit}
g_{\mu\nu}= \bar g_{\mu\rho} ( e^{\gb^{-1}h} )^\rho{}_\nu\ .
\ee
This choice ensures that $g_{\mu\nu}$ and $\bar g_{\mu\nu}$ have the same signature, even if the fluctuations are large. Throughout the work quantities constructed from the background metric will be denoted with a bar.

The evaluation of the flow equation for the ansatz \eqref{Gammaans} can be significantly simplified by choosing a suitable background. Since our ansatz projects the full RG flow onto functions of the Ricci scalar, it is convenient to work with $\bar g_{\mu\nu}$ being a one-parameter family of metrics on the maximally symmetric $d$-sphere with (arbitrary) radius $a$. In this case the Riemann tensor and the Ricci tensor are determined by $\Rb$,
\be\label{max:sym}
\bar R _{\mu\rho\nu\sigma} = \frac {\bar R}{d(d-1)}
\left( \bar g_{\mu\nu}\bar g_{\rho\sigma} - \bar g_{\mu\sigma }
\bar g_{\nu\rho} \right) \, ,  \qquad 
\bar R _{\mu \nu} = \frac {\bar R}{d} \, \bar g_{\mu\nu} \, , 
\ee
and the Ricci scalar is constant, $\bar D_\mu \bar R=0$. Furthermore, the scalar curvature is related to the radius $a$ of the sphere,
\be
\bar R =  \frac 1 {a^2} d (d-1) \, , \label{curv}
\ee
and the volume $V_d$ of the background is given by
\be\label{volsphere}
V_d = \frac{2 \, \pi^{(d+1)/2}}{\Gamma((d+1)/2)} \, a^d =  \frac{2 \, \pi^{(d+1)/2}}{\Gamma((d+1)/2)} \, \left( \frac{d (d-1)}{\Rb} \right)^{d/2} \, . 
\ee
Defining the Laplacian $\Delta = -\gb^{\mu\nu} \Db_\mu \Db_\nu$, the complete set of eigenvalues $\lambda_\ell^{(s)}$ together with their degeneracies $M_\ell^{(s)}$ for $\Delta$ acting on irreducible spin representations have been determined in \cite{Rubin:1984tc,Camporesi:1995fb}. This data is collected in Table \ref{Tab.ev}. In particular, the spectrum of the Laplacian acting on spinor fields is obtained from the eigenvalues of $-\bar \Dslash~^2$ \cite{Camporesi:1995fb},
\be
\lambda_l^{-\bar {\DslashIndex}~^2} = \frac{1}{a^2} \left(l + \tfrac{d}{2} \right)^2
\; , \qquad 
M_l^{-\bar \DslashIndex~^2} = 2^{\lfloor d/2+1 \rfloor}\frac{(\ell+d-1)!}{(d-1)! \, \ell !} \, , \qquad \ell = 0,1,\ldots \, , 
\ee
in combination with the Lichnerowicz formula $-\bar \Dslash~^2 = \Delta + \Rb/4$. Here $\lfloor \ldots \rfloor$ denotes the floor function.
\begin{table}[t!]
	\begin{tabular}{c|ccc}
	\; \;	spin $s$ \; \; & $\lambda_\ell^{(s)}$ & $M_\ell^{(s)}$ & \\ \hline \hline
		$0$ & \qquad $\frac 1 {a^2} \ell (\ell+d-1)$ \quad \;  & \qquad $\frac{(\ell+d-2)!}{(d-1)! \, \ell !}(2\ell +d-1)$ \quad  \; & \qquad $\ell = 0,1, \ldots$ \quad \; \\[1.1ex]
		$\tfrac{1}{2}$ & $\frac 1 {a^2} (\ell^2 + d\ell + \tfrac{d}{4})$ & $ 2^{\lfloor d/2+1 \rfloor}\frac{(\ell+d-1)!}{(d-1)! \, \ell !}$ & \qquad $\ell = 0,1, \ldots$ \quad \; \\[1.1ex]
		$1$ & $\frac 1 {a^2} \left(\ell (\ell+d-1) -1 \right)$ & 
		$\frac{(\ell+d-3)!}{(d-2)! (\ell +1)!}(2\ell +d-1)(\ell+d-1)\ell$ & $\ell = 1,2,\ldots$ \\[1.1ex]
		$2$ &  \; \; $\frac 1 {a^2} \left(\ell (\ell+d-1) -2 \right)$  \; \;  &  \; \;  
		$\tfrac{(d+1)(d-2)(l+d)(l-1)(2l+d-1)(l+d-3)!}{2(d-1)!(l+1)!}$  \; \; 
		& $\ell = 2,3,\ldots$ \\ \hline \hline
	\end{tabular}
	\caption{\label{Tab.ev} Eigenvalues $\lambda_\ell^{(s)}$ and their degeneracy $M_\ell^{(s)}$ for the Laplacian $\Delta = -\gb^{\mu\nu} \Db_\mu \Db_\nu$ acting on
		scalars ($s=0$), Dirac fermions ($s=1/2$), transverse vectors ($s=1$) and transverse-traceless matrices ($s=2$). The bosonic results are taken from \cite{Rubin:1984tc} while the fermionic case has been derived in \cite{Camporesi:1995fb}.}
\end{table}

Finally, the computation is simplified by decomposing the fluctuation fields into their irreducible spin components, see \cite{Lauscher:2001ya} for an extended discussion. For the gravitational fluctuations  this is achieved by the York 
decomposition \cite{York:1973ia}
\be
h_{\mu\nu} = h^{TT}_{\mu\nu} + \bar D_\mu\xi_\nu + \bar D_\nu\xi_\mu +
(\bar D_\mu \bar D_\nu  -\frac{1}{d} \bar g_{\mu\nu} \bar D^2 )\sigma +
\frac{1}{d} \bar g_{\mu\nu} h,
\label{york}
\ee
which expresses $h_{\mu\nu}$ in terms of a transverse-traceless tensor $h^{TT}_{\mu\nu}$ (spin $s=2$), a transverse vector $\xi_\nu$ (spin $s=1$) and two scalar fields $\sigma,h$ subject to the differential constraints
\be
\bar D^\mu h^{TT}_{\mu\nu} = 0 \, , \quad \bar g^{\mu\nu} h^{TT}_{\mu\nu} =0 \, , \quad
\bar D^\mu \xi_\mu=0.
\ee
Similarly, a vector field $A_\mu$ is decomposed into a transverse vector $A_\mu^T$ and a scalar $a$ according to
\be\label{Tdec}
A_\mu = A_\mu^T + \bar{D}_\mu \, a \, , \qquad \Db^\mu A_\mu^T = 0 \, .  
\ee
Notably, not all eigenmodes of the Laplacian contribute to the decompositions \eqref{york} and \eqref{Tdec}. A constant mode $a$ drops out of the transverse decomposition \eqref{Tdec} while in the York decomposition the two lowest eigenmodes of $\sigma$ and the lowest vector mode $\xi_\mu$ of the Laplacian do not change the right-hand-side of \eqref{york}. These zero modes must then be removed by hand in order to make the decompositions into irreducible spin components bijective. Moreover, the decompositions give rise to operator-valued Jacobians. On a spherical background these are given by
\be\label{fieldredef}
\cJ^{\rm vec} = \Det_{(1)}\left(\Delta - \tfrac{\Rb}{d}\right)^{1/2} \, , \quad
\cJ^{\sigma} = \Det_{(0)}\left(\Delta^2 - \tfrac{\Rb}{d-1} \Delta\right)^{1/2} \, , \quad 
\cJ^{a} = \Det_{(0)}\left(\Delta\right)^{1/2} \, . 
\ee
\subsection{Trace contributions from the gravitational and matter sector}
\label{sec:RGgrav}
Given the ansatz \eqref{Gammaans} the flow of $\Gamma_k$ will be sourced by quantum fluctuations in the gravitational and matter sector
\be\label{deftraces}
\partial_t \Gamma_k = T^{\rm grav} + T^{\rm matter} \, . 
\ee
The construction of the gravitational sector follows \cite{Ohta:2015efa,Ohta:2015fcu}. In this setting, $\Gamma_k^{\rm grav}$
is supplemented by a classical gauge-fixing term
\be
\label{gf1}
\Gamma^{\rm gf}=\frac{1}{2\alpha}\int d^d x \sqrt{\bar g}\,\bar g^{\mu\nu}F_\mu F_\nu \, , \qquad F_\mu = \bar D_\rho h^\rho{}_\mu-\frac{\beta +1}{d} \bar D_\mu h\, .
\ee
Expressing $F_\mu$ in terms of the component fields \eqref{york} one has
\bea
F_\mu &=& -\left( \Delta   -\tfrac{\bar R}{d} \right)\xi_\mu
- \tfrac{1}{d} \bar D_\mu
\left(\left[(d-1)\Delta - \Rb \right]\sigma+ \beta \, h\right).
\label{gf}
\eea
Following \cite{Benedetti:2011ct} this suggests to recast the scalar fields in terms of a gauge-invariant field $s$ and a gauge-dependent degree of freedom $\chi$
\be\label{scalarsector}
s= h + \Delta \sigma \, , \qquad \chi = \frac{[(d-1)\Delta - \Rb] \sigma + \beta h}{(d-1-\beta)\Delta - \Rb}
\ee
where the denominator in $\chi$ is fixed by requiring that the transformation has a constant Jacobian. Expressing the gauge-fixing term in terms of these fields leads to
\be
\Gamma^{\rm gf} = \frac{1}{2\alpha}\int d^d x \sqrt{\bar g} \, \Big\{
\xi^\mu \left[  \Delta - \tfrac{\Rb}{d} \right]^2 \xi_\mu + \tfrac{(d-1-\beta)^2}{d^2} \chi \, \left[ \Delta \left( \Delta - \tfrac{\Rb}{(d-1-\beta)} \right)^2 \right] \,\chi
\Big\} \, . 
\ee

The ghost action associated with the gauge fixing \eqref{gf1} is obtained in the standard way. Restricting to terms quadratic in the fluctuation fields, it reads
\bea
\Gamma^{\rm ghost}&=&\int d^dx\sqrt{\bar g} \, \, \bar C^\mu\left[ \,
\delta_\mu^\nu \, \bar D^2 
+\left(1-2 \, \tfrac{\beta+1}{d}\right) \bar D_\mu \bar D^\nu+\tfrac {\bar R} d \, \delta_\mu^\nu\right]C_\nu .
\eea
Decomposing the ghosts into its transversal and longitudinal part, $C_\nu = C_\nu^T + \bar D_\nu  \, C^{L}$, 
one obtains
\bea
\Gamma^{\rm ghost} = - \int d^dx\sqrt{\bar g} \, \,  \left\{   \bar C^{T \mu} \left[ \Delta  - \tfrac {\bar R}{d}  
\right]C^T_\mu + 2 \, \tfrac{d-1-\beta}d \, \bar C^{L} \left[  \Delta - 
\tfrac {\bar R}{d-1-\beta} \right] \, \Delta \, C^{L} \right\} .
\eea

The gravitational sector is completed by the expansion of $\Gamma^{\rm grav}_k[g] = \Gamma^{\rm grav}_k[\gb] + \cO(h) + \Gamma_k^{\rm quad}[h;\gb]+\ldots$. For the exponential split \eqref{ExpSplit} the terms quadratic in the fluctuation fields are given by \cite{Ohta:2015fcu} 
\bea
\Gamma_k^{\rm quad} &=& \int d^dx \sqrt{\gb} \,  \Big\{
-\tfrac{1}{4} \, f'(\bar R) \, h_{\mu\nu}^{TT} \, \Big[\Delta +\tfrac{2}{d(d-1)}\bar R \Big]\,  h^{TT\, \mu\nu}
\nonumber \\ && 
+ \tfrac{d-1}{4d} \, s \, \Bigg[ \tfrac{2(d-1)}{d} \, f''(\bar R ) \, \Big(\Delta -\tfrac{\bar R}{d-1}\Big)
+ \tfrac{d-2}{d}f'(\bar R )\Big] \Big[ \Delta-\tfrac{\bar R}{d-1} \Big] \, s
\nonumber \\  && 
+ h \, \Big[ \tfrac{1}{8} f(\bar R )-\tfrac{1}{4d} \bar  R \, f'(\bar R) \Big] \, h \Big\} \, .
\label{HessianGrav}
\eea
Note that all terms containing the spin-1 component $\xi_\mu$ canceled out. 

At this stage, it is useful to collect the determinants arising from the various spin sectors. The transverse vector sector receives contributions from the Jacobian in the transverse-traceless decomposition \eqref{fieldredef}, from the transverse ghosts, and from $\xi^\mu$ in the gauge-fixing term. All one-loop determinants have the same form, such that they combine according to
\be\label{vecdet}
\Det_{(1)}\left(\Delta - \tfrac{\Rb}{d}\right)^{1/2} \, \Det_{(1)}\left(\Delta - \tfrac{\Rb}{d}\right) \, \Det_{(1)}\left(\Delta - \tfrac{\Rb}{d}\right)^{-1} = \Det_{(1)}\left(\Delta - \tfrac{\Rb}{d}\right)^{1/2} \, . 
\ee
In the scalar sector, one combines the contributions from $\chi$, the longitudinal ghost $C^L$, and the scalar determinants from the field decompositions in the transverse-traceless and ghost decomposition
\be
\begin{split}
& \Det_{(0)}\left(\Delta\right)^{-1/2} \,
\Det_{(0)}\left(\Delta - \tfrac{\Rb}{d-1-\beta} \right)^{-1} \cdot  \Det_{(0)}\left(\Delta\right) \, 
  \Det_{(0)}\left(\Delta - \tfrac{\Rb}{d-1-\beta} \right) \cdot \\ &
  \Det_{(0)}\left(\Delta\right)^{1/2} \,
   \Det_{(0)}
  \left(\Delta - \tfrac{\Rb}{d-1}\right)^{1/2} \cdot \Det_{(0)}\left(\Delta\right)^{-1} = 
  \Det_{(0)}\left(\Delta - \tfrac{\Rb}{d-1}\right)^{1/2} \, ,
  \end{split}
\ee
where we used the $\cdot$ to separate the contributions from the various sectors. Note that the remaining scalar determinant can be absorbed by evoking the field redefinition $s \rightarrow \tilde s = \left[\Delta - \tfrac{\Rb}{d-1} \right]^{1/2} s$, which simplifies the contribution of the scalar sector in \eqref{HessianGrav}. The cancellation of the scalar determinants is actually \emph{independent} of the choice of the gauge-fixing parameter $\beta$. It solely relies on the field redefinition \eqref{scalarsector} used to disentangle the gauge-invariant and gauge-dependent field contributions.

The structure of the Hessians is further simplified by adopting ``physical gauge'' $\beta \rightarrow - \infty$, $\alpha \rightarrow 0$. From \eqref{scalarsector} one finds that the limit $\beta \rightarrow - \infty$ aligns $\chi$ and $h$ such that $\chi \propto h$. Subsequently evoking the Landau limit $\alpha \rightarrow 0$ then ensures that the $h^2$ term appearing in $\Gamma_k^{\rm quad}$ does not contribute to the flow equation. In this way the contributions of the fields in the gravitational sector is maximally decoupled: $\Gamma_k^{\rm quad}$ gives the contributions for $h^{\rm TT}_{\mu\nu}$ and $s$, while the gauge-fixing term determines the quantum fluctuations of the transverse vector $\xi_\mu$ and scalar $\chi$.

 The final ingredient in writing down the flow equation \eqref{FRGE} is the regulator $\cR_k(\Box)$. Since one of the main objectives of this work is to understand the role of different coarse-graining operators, we introduce the operators
\be
\Box_{S,D,V,T}^{G,M} \equiv \Delta - \alpha_{S,D,V,T}^{G,M} \bar R \, ,
\ee 
which, besides the Laplacian, also contain an endomorphism parameter $\alpha_{S,D,V,T}^{G,M}$. Here the superscript indicates if the operator belongs to the gravitational (G) or matter sector (M) while the subscript gives the spin of the corresponding fields. In case of ambiguities, we will add additional numbers to the spin index. We then define the regulator $\cR_k(\Box)$ through the replacement rule
\be\label{TypeIrep}
\Box \mapsto P_k(\Box) \equiv \Box + R_k(\Box) \, ,
\ee
where it is understood that the endomorphism parameters contained in the coarse graining operators $\Box$ may differ for different fields.

Based on \eqref{HessianGrav} and \eqref{vecdet} we have now have all ingredients for writing down the gravitational contribution to the flow of $f_k(R)$. The gravitational sector gives rise to three contributions associated with the transverse-traceless fluctuations $h_{\mu\nu}^{TT}$, the gauge-invariant scalar $s$ and the vector determinant \eqref{vecdet}:
\be
T^{\rm grav} = T^{\rm TT} + T^{\rm ghost} + T^{\rm sinv} \, . 
\ee
The explicit expressions for the traces are given by
\begin{subequations}\label{gravitytraces}
	\begin{align}
	T^{\rm TT} = &  \, \frac{1}{2}\Tr_{(2)}  \left[ 
	\left(f'(\bar R)( P_k^T
		+\alpha_T^G \bar R+\tfrac{2}{d(d-1)}\bar R)\right)^{-1}
	\partial_t \left( f'_k(\bar R)  R_k^T \right) \right] \, , \\
T^{\rm sinv} = & \, 
\frac 1 2 \Tr^{\prime\prime}_{(0)} \left[ \left(f''_k(\bar R) (P_k^S
+\alpha_S^G\bar R-\tfrac{1}{d-1}\bar R) +\tfrac{d-2}{2(d-1)}f'_k(\bar R)\right)^{-1}  \partial_t \left( f''_k(\bar R) R_k^S \right)  \right] \, ,
\\
T^{\rm ghost} = - & \,   \frac 1 2 \Tr^{\prime}_{(1)} \left[ 
\left(P_k^V +\alpha_V^G \bar R -\tfrac{1}{d}\bar R\right)^{-1} \partial_t R_k^V
\right]  \, . 
	\end{align}
\end{subequations}
Here the number of primes on the traces indicate the number of modes which have to be discarded. The subscript on the traces, on the other hand, specify the spin of the fields. By construction, the result agrees with \cite{Ohta:2015fcu}.

The contribution of the minimally coupled matter fields \eqref{mattersector} to the gravitational flow can be constructed along the same lines as in the gravitational sector: one first decomposes the vector field into its transverse and longitudinal parts according to \eqref{Tdec}, computes the Hessians $\Gamma^{(2)}$, and determines the regulator function according to the prescription \eqref{TypeIrep}. The resulting contribution is given by
\be
T^{\rm matter} = T^{\rm scalar} + T^{\rm Dirac}  + T^{\rm vector} 
\ee
where
\begin{subequations}\label{mattertraces}
	\begin{align}
	T^{\rm scalar} = &  \, \frac{N_S}{2} \, \Tr_{(0)}  \left[ (P_k^S+\alpha_S^M \bar R)^{-1} \, \partial_t R_k^S \right] \, , \\
	T^{\rm Dirac} = & \, - \frac{N_D}{2}  \Tr_{(1/2)} \left[ (P_k^D+\alpha_D^M \bar R + \tfrac{1}{4} \Rb)^{-1} \, \partial_t R_k^D \right] \, , \\ 
	T^{\rm vector} = & \, \frac{N_V}{2}  \Tr_{(1)} \left[ (P_k^{V_1}+\alpha_{V_1}^M \bar R + \tfrac{1}{d} \Rb)^{-1} \, \partial_t R_k^{V_1} \right] +
	\frac{N_V}{2}  \Tr_{(0)}^\prime \left[ (P_k^{V_2}+\alpha_{V_2}^M \bar R)^{-1} \, \partial_t R_k^{V_2} \right] \\ \nonumber & \,
	- N_V  \Tr_{(0)}^\prime \left[ (P_k^{V_2}+\alpha_{V_2}^M \bar R)^{-1} \, \partial_t R_k^{V_2} \right].
    \end{align}
\end{subequations}
The three traces in $T^{\rm vector}$ capture the contribution from the transverse vector field, the longitudinal modes, and ghost fields, respectively. Again the number of primes indicates that the corresponding number of lowest eigenmodes should be removed from the trace. In addition, we have equipped each sector with its own endomorphism parameter $\alpha$. Following \cite{Ohta:2015efa,Ohta:2015fcu}, we also insisted on a ``mode-by-mode'' cancellation between the matter and ghost modes, so that the corresponding traces come with the same number of primes and endomorphism parameter. The full, projected flow equation is then obtained by substituting \eqref{gravitytraces} and \eqref{mattertraces} into \eqref{deftraces}.

\subsection{Constraining the coarse-graining operator}
\label{sec:endomorphism}
Notably, the values of the endomorphism parameters $\alpha$ may not be chosen arbitrarily. On physical grounds one requires that
\begin{enumerate}
	\item[1.] For any fluctuation contributing to the operator traces in the flow equation the argument of the regulator $\cR_k(\Box)$, $\Box = \Delta - \alpha \Rb$, should be positive-semidefinite.
\end{enumerate}
and
\begin{enumerate}
	\item[2.] The denominators appearing in the trace-arguments should be free of poles on the support of $\Box$. In other words, the ``mass-type'' terms provided by the background curvature should not correspond to a negative squared-mass.
\end{enumerate}
At first sight the second condition may seem somewhat less compelling since these types of singularities are removed when the flow equation is expanded in powers of the background curvature. Taking into account that the approximate solutions of the flow equation arising from this procedure should ultimately have an extension to solutions of the full flow equations, constraining the coarse graining operator to those which do not give rise to such extra singularities is a sensible requirement.

Practically, the first condition translates into the requirement that $\alpha \Rb$ must be smaller than the lowest eigenvalue contributing to a given trace. Taking into account the omitted lowest eigenmodes (indicated by the primes in  \eqref{gravitytraces}) the resulting constraints in the gravitational sector are
\be\label{eqc1g}
\alpha_T^G \le \frac 2{d-1} \, , \qquad 
\alpha_S^G \le \frac{2(d+1)}{d(d-1)} \, , \qquad 
\alpha_V^G \le \frac{2d+1}{d(d-1)} 
 \, .
\ee
Analogously, the endomorphism parameters in the matter sector should satisfy
\be\label{eqc1m}
\alpha_S^M \le 0 \, , \qquad 
\alpha^M_D \le \frac{1}{4(d-1)} \, , \qquad
\alpha^M_{V_1} \le \frac{1}{d} \, , \qquad
\alpha^M_{V_2} \le \frac{1}{d-1} \, . 
\ee
Here different bounds for fields with the same spin arise due to a different number of fluctuation modes excluded from the traces.

The second condition is evaluated by replacing $P_k \rightarrow k^2$ and subsequently writing the propagators in terms of the dimensionless curvature $r \equiv \Rb k^{-2}$. The denominators then take the form $(1+(c_d+\alpha)r)$ where the constants $c_d$ depend on the trace under consideration and can be read off from \eqref{gravitytraces} and \eqref{mattertraces}. For fixed background curvature $\Rb$ and $k \in [0,\infty[$ the dimensionless curvature takes values on the entire positive real axis $r \in [0,\infty[$. The absence of poles results in the condition $\alpha \ge - c_d$. In the gravitational sector this entails
\be\label{eqc2g}
\alpha_T^G \ge -\frac 2{d(d-1)} \, , \qquad 
\alpha_S^G \ge \frac{1}{d-1} \, , \qquad 
\alpha_V^G \ge \frac{1}{d} 
\, ,
\ee
while for the matter fields the bounds are
\be\label{eqc2m}
\alpha_S^M \ge 0 \, , \qquad 
\alpha^M_D \ge - \frac{1}{4} \, , \qquad
\alpha^M_{V_1} \ge - \frac{1}{d} \, , \qquad
\alpha^M_{V_2} \ge 0 \, . 
\ee
The bound on $\alpha_S^G$ reported in \eqref{eqc2g} may be less stringent though, since the quoted value does not take into account possible contributions from the function $f_k(\Rb)$ which can only be computed at the level of solutions. Notably, both sets of conditions \eqref{eqc1g}, \eqref{eqc1m} and \eqref{eqc2g},\eqref{eqc2m} can be met simultaneously. This requires non-zero endomorphism parameters $\alpha_S^G$ and $\alpha_V^G$ though.

We close our discussion by introducing two widely used choices for the coarse graining operators termed ``Type I'' and ``Type II'' (see \cite{Codello:2008vh} for a detailed discussion). In this case the endomorphism parameters are chosen as
\begin{subequations}\label{endomorphism}
	\begin{align}\label{endtypeI}
	& \mbox{Type I:}  \quad \alpha^G_T = \alpha^G_S = \alpha^G_V  =  \alpha^M_D = \alpha^M_{V_1} = \alpha^M_{V_2} = \alpha^M_S = 0 \, , \\ \label{endtypeII}
		& \mbox{Type II:} \quad
		\alpha^G_T = - \tfrac{2}{d(d-1)} \, , \; 
		\alpha^G_S = \tfrac{1}{d-1} \, , \; 
		\alpha^G_V = \tfrac{1}{d} \, , \; 
		\alpha^M_D = - \tfrac{1}{4} \, , \; 
		\alpha^M_{V_1} = - \tfrac{1}{d} \, , \;
		\alpha^M_{V_2} = \alpha^M_S = 0 \, .
	\end{align}
\end{subequations}
For the Type I choice the coarse graining operator $\Box$ agrees with the Laplacian acting on the corresponding spin fields. The Type II coarse graining operator is tailored in such a way that it removes the scalar curvature from the propagators.\footnote{The use of a Type I and Type II coarse graining operator should not be confused with a ``change of the regulator function''. Changing the value of the $\alpha$'s presumably results in quantizing a different theory \cite{us1}. One way to fix the values of $\alpha$ is provided by the principle of equal lowest eigenvalues \cite{Demmel:2014hla}, but at this stage we treat the endomorphisms as free parameters.} By construction it satisfies both condition 1 and 2.

In order to trace the dependence of the RG flow on the choice of coarse graining operator, we furthermore introduce an ``interpolating'' coarse graining operator where
\be\label{regtypei}
\mbox{Type i:} \quad
\alpha^G_T = - \tfrac{2c}{d(d-1)} \, , \; 
\alpha^G_S = \tfrac{c}{d-1} \, , \; 
\alpha^G_V = \tfrac{c}{d} \, , \; 
\alpha^M_D = - \tfrac{c}{4} \, , \; 
\alpha^M_{V_1} = - \tfrac{c}{d} \, , \;
\alpha^M_{V_2} = \alpha^M_S = 0 \, .
\ee
This one-parameter family contains one free parameter $c$ and interpolates continuously between a coarse graining operator of Type I for $c=0$ and Type II for $c=1$. In particular, this construction will be very useful in order to understand the fixed point structure of gravity-matter systems in Sect.\ \ref{sec:fixedpoints}.
\section{Evaluating operator traces as spectral sums}
\label{sec:spectrum}
The next step consists in explicitly evaluating the traces \eqref{gravitytraces} and \eqref{mattertraces} and rewrite them as explicit functions of the scalar curvature. The main result is the partial differential equation \eqref{pdfgeneral} and its restriction to four dimensions \eqref{pdf4d} which governs the scale-dependence of $f_k(R)$ in the presence of minimally coupled matter fields.  

 Our computation follows the strategy \cite{Reuter:2008qx,Benedetti:2012dx,Benedetti:2013jk,Demmel:2014sga,Ohta:2015efa,Ohta:2015fcu} and performs the traces as sums over eigenvalues of the corresponding Laplacians. Furthermore, we employ a Litim-type regulator, setting
\be\label{RLitim}
R_k(z) = (k^2-z) \theta(k^2 - z) \, , \qquad \partial_t R_k(z) = 2 k^2 \theta(k^2 - z) \, . 
\ee
For finite $k$, the presence of the step-function in the regulator entails
that only a finite number of eigenvalues contribute to the mode sum. Moreover the propagators are independent of $\Delta$ and can be pulled out of the sums. As a consequence the traces reduce to finite sums
over the degeneracies of the eigenvalues, possibly weighted by the corresponding eigenvalue. These sums take the form
\be\label{sumdeg}
S^{(s)}_{d}(N) \equiv \sum_{\ell = \ell_{\rm min}}^{N} \, M_\ell^{(s)} \, , \qquad
\widetilde{S}^{(s)}_{d}(N) \equiv \sum_{\ell = \ell_{\rm min}}^{N} \, \lambda_\ell^{(s)} \, M_\ell^{(s)} \, ,
\ee
were $N$ is a (finite) integer determined by the regulator, and the eigenvalues and degeneracies are listed in Table \ref{Tab.ev}. In the matter sector, all traces have the structure $S^{(s)}_d(N)$ while the gravitational sector gives rise to both types of contributions. The occurrence of contributions of the form $\widetilde{S}^{(s)}_{d}(N)$ can be traced back to the presence of scale-dependent coupling constants in the regulator functions which only occurs in the gravitational sector and are absent in the matter traces. In this section we use the sums \eqref{sumdeg} to explicitly evaluate the right-hand-side of the flow equation by summing over the eigenvalues of the differential operators. 

Carrying out the sums for scalars ($s=0$), Dirac fermions ($s=1/2$), 
and transverse vectors $(s = 1)$, and transverse-traceless tensors ($s=2$) results in
\begin{subequations}\label{mattersums}
	\begin{align}
	S^{(0)}_{d}(N) = & \, \left(2N+d\right) \frac{(N+d-1)!}{d! \, N!} \, ,	\\
	S^{(1/2)}_{d}(N) = & \, 2^{\lfloor d/2+1 \rfloor} \, \frac{(N+d)!}{d! \, N!} \, , \\
	S^{(1)}_{d}(N) = & \, 1 + \frac{d-1}{d!} \, \frac{(2N+d) \, (N^2+dN-1)\, (N+d-2)!}{(N+1)!} \, , \\
	S^{(2)}_{d}(N) = & \,\tfrac{(d+2)(d+1)}{2} + \tfrac{(d+1)(2N+d)\big((d-2) (N^2+dN) - (d+2)(d-1)\big) (N+d-2)!}{2 \, d! \, (N+1)!} \, .  
	\end{align}
\end{subequations}
These results may readily be confirmed by applying proof by induction techniques. All expressions are polynomials of order $d$ in $N$. The sums weighted by the eigenvalues can be performed in the same way. In this case it suffices to consider the cases $s=0$ and $s=2$, yielding
\begin{subequations}\label{gravitonsum}
	\begin{align}
	\widetilde{S}^{(0)}_{d}(N) = & \, \frac{ (2N + d) (N + d)!}{a^2 \, (d+2) \, (d-1)! (N-1)!} \, , \\
	%
	%
	\widetilde{S}^{(2)}_{d}(N) = & \,
	- \tfrac{2d(d+1)}{a^2}
	+ \tfrac{(d+1) (2N+d) \big((d-2)(N^4 + 2dN^3 +(d^2-d-5)N^2 -d(d+5)N) + 4 (d-1) (2 + d) \big) (N+d-2)!}{2 a^2 (2 + d) ( d-1)! (N+1)!} \, . 
	\end{align}
\end{subequations}
These expressions are again polynomials in $N$ with order $d+2$. The increased order thereby compensates the factor $a^2$ such that both \eqref{mattersums}
and $\eqref{gravitonsum}$ exhibit the same scaling behavior for as $R \rightarrow 0$.

The value $N$ at which the sums are cut off is given by the largest integer $N^{(s)}_{\rm max}$ satisfying the inequality 
$\lambda^{(s)}_{N^{(s)}_{\rm max}} - \alpha \Rb \le k^2$. 
Substituting the eigenvalues listed in Table \ref{Tab.ev} and solving this condition for $N^{(s)}_{\rm max}$ yields
\begin{subequations}\label{Ncutoff}
	\begin{align}
	N^{(0)}_{\rm max} = & \, - \tfrac{d-1}{2} - p^{(0)} + \tfrac{1}{2} \sqrt{d^2-2d+1 + 4 d (d-1) \left(\tfrac{1}{r} + \alpha\right) + q^{(0)}} \, , \\
	N^{(1/2)}_{\rm max} = & \, - \tfrac{d}{2} - p^{(1/2)} + \sqrt{ d (d-1)  \left( \tfrac{1}{r} + \tfrac{1}{4} +  \alpha\right) + \tfrac{1}{4} \,  q^{(1/2)} } \, , \\
	N^{(1)}_{\rm max} = & \, - \tfrac{d-1}{2} -  p^{(1)} + \tfrac{1}{2} \sqrt{d^2-2d+5 + 4 d (d-1) \left(\tfrac{1}{r} + \alpha\right) + q^{(1)} } \, , \\
	N^{(2)}_{\rm max} = & \, - \tfrac{d-1}{2} -  p^{(2)} + \tfrac{1}{2} \sqrt{d^2-2d+9 + 4 d (d-1) \left(\tfrac{1}{r} + \alpha\right) + q^{(2)} } \, ,
	\end{align}
\end{subequations}
\begin{figure}[t!]
	\includegraphics[width=0.48\textwidth]{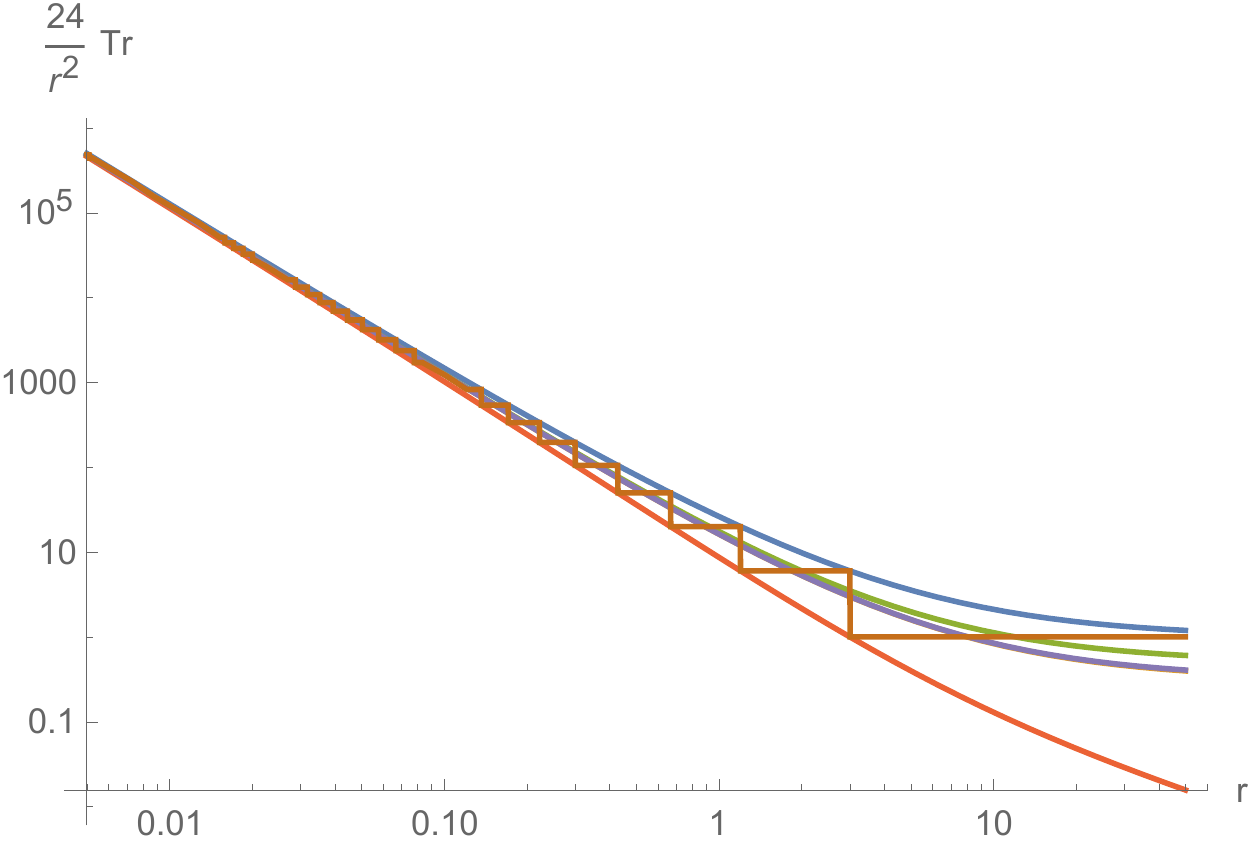}
	\includegraphics[width=0.48\textwidth]{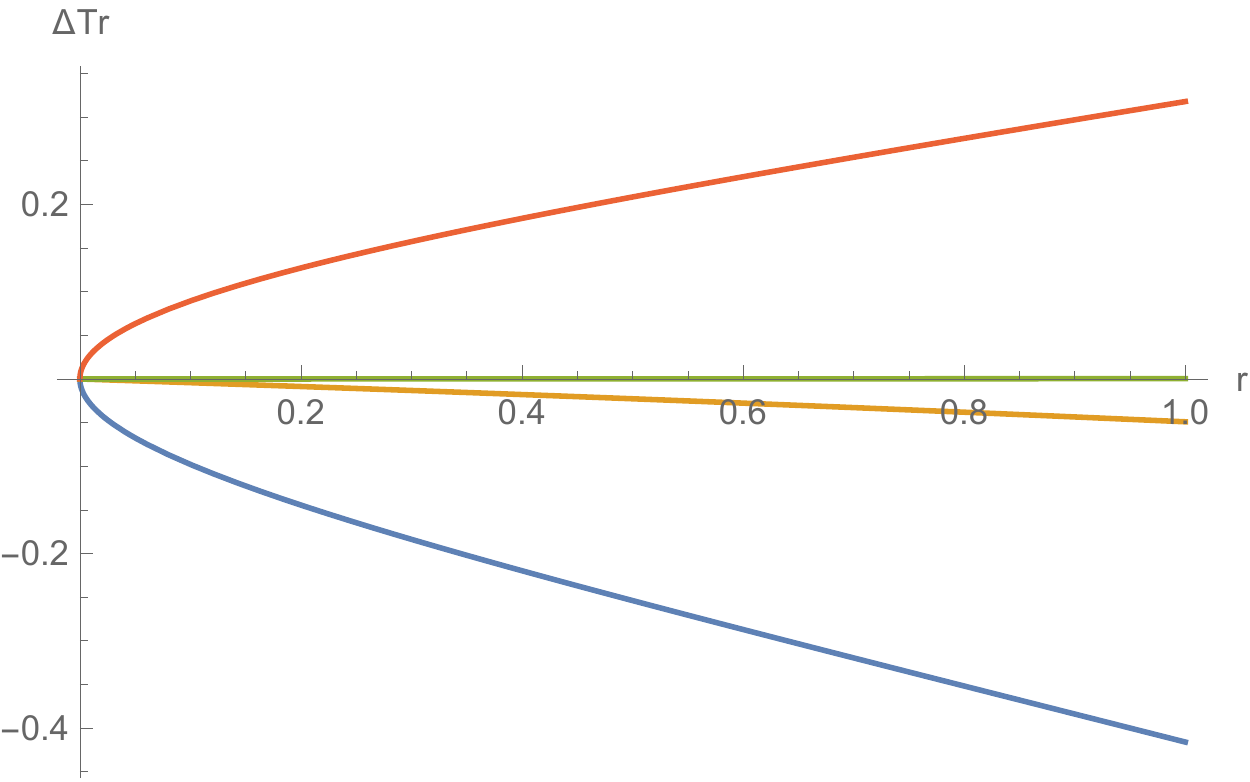}
	\caption{Comparison of smoothing procedures applied to the four-dimensional scalar trace without endomorphism, $\Tr_{(0)}[\theta(k^2-\Delta)]$, as a function of the dimensionless curvature $r=\Rb/k^2$. The left panel depicts the staircase behavior of the sum over modes (horizontal lines). The upper (lower) staircase approximations interpolate between the upper (lower) points of the discrete result. The early-time expansion of the heat kernel, as well as the averaged ($q^{(s)}=0$) and optimized averaged  ($q^{(s)}=-2$) interpolations lie between these extreme curves. The right diagram displays the difference $\Delta\Tr$ obtained from evaluating the scalar trace with the early-time expansion of the heat kernel (reference) and (from bottom to top) the upper staircase, averaged, optimized averaged,  and the lower staircase interpolation. On the shown interval the relative difference between the optimized averaged interpolation and the early-time expansion of the heat kernel \eqref{heatbench} is smaller than $6 \times 10^{-4}$. \label{fig.smoothing}}
\end{figure}
with $p^{(s)} = q^{(s)} = 0$ and $r\equiv\Rb/k^2$. The sums then extend up to the integer part of these bounds which results in a discontinuous structure in the flow equation. This is illustrated in the left panel of Fig.\ \ref{fig.smoothing}. Since for any fixed dimension $d$ the expressions \eqref{mattersums} and \eqref{gravitonsum} reduce to polynomials in $N$, one may substitute the corresponding thresholds \eqref{Ncutoff} and treat the resulting expressions as being continuous in the dimensionless curvature $r$. For $p^{(s)} = q^{(s)} = 0$ this results in the upper staircase interpolation shown as the top curve in the left diagram of  Fig.\ \ref{fig.smoothing}. The lower staircase curve (connecting the lower points of the discontinuous steps) is obtained from setting $p^{(s)} = -1$ and $q^{(s)} = 0$. The  interpolation used in \cite{Ohta:2015efa,Ohta:2015fcu} averages the sums \eqref{sumdeg} evaluated at  $N_{\rm max}^{(s)}$ and $N_{\rm max}^{(s)}-1$ setting $p^{(s)} = q^{(s)} = 0$. This procedure removes the non-analytic terms from the sums
once the volume $V_d$ has been factored out. We will employ this averaging interpolation in the sequel. It is then convenient to define sums tailored to the averaged interpolation, setting
\be\label{averagedinterpolationsums}
T^{(s)}_d(N) \equiv \tfrac{1}{2} \left( S^{(s)}_d(N) + S^{(s)}_d(N-1) \right) \, , \qquad \widetilde{T}_d(N) \equiv \tfrac{1}{2} \left( \widetilde{S}^{(s)}_d(N) + \widetilde{S}^{(s)}_d(N-1) \right) \, . 
\ee

At this stage, it is instructive to compare the evaluation of the operator traces in terms of spectral sums to the results obtained from the early-time expansion of the heat kernel. Applying standard Mellin-transform techniques \cite{Reuter:1996cp,Codello:2008vh} one has
\be\label{trheatkernel}
\Tr_{(s)} \theta(k^2 - \Delta) = \frac{k^d}{(4 \pi)^{d/2}} \, V_d \, f_{(s)}(r;d)
\ee
where
\be\label{frd}
f_{(s)}(r;d) = \left( \frac{{\rm tr} \, {\bf a}_0^{(s)}}{\Gamma(d/2+1)} + \frac{{\rm tr} \, {\bf a}_2^{(s)}}{ \Gamma(d/2)} \, k^{-2} \right) + \mathcal{O}(r^2)
\ee
and the coefficients ${\bf a}_n^{(s)}$ can be found in \cite{Lauscher:2001ya}. In particular, for a scalar field in $d=4$ dimensions the early-time expansion of \eqref{trheatkernel} gives
\be\label{heatbench}
f_{(0)}(r;4) = \tfrac{1}{2} + \tfrac{1}{6}r + \tfrac{29}{2160} r^2 \, .  
\ee
Generically, evaluating \eqref{mattersums} for averaging interpolation reproduces the leading term in \eqref{frd} while the subleading coefficient multiplying $r$ will match for specific values $q^{(s)} \not = 0$, only. This suggests an optimized averaged interpolation function where
\be\label{qopt}
q^{(s)} = -\tfrac{2}{3} (d-1) \, . 
\ee
The results obtained from the various interpolations are then compared in the right panel of Fig.\ \ref{fig.smoothing} which displays the difference of the functions $f_{(0)}(r,4)$ obtained in \eqref{heatbench} and (from bottom to top) the upper-staircase interpolation, the optimized interpolation based on the values \eqref{qopt}, the averaged interpolation, and the lower-staircase interpolation. As expected the optimized interpolation leads to an expansion which gives the best approximation to the early-time heat kernel.

Based on the mode sums \eqref{mattersums} and \eqref{gravitonsum} together with the cutoffs \eqref{Ncutoff} it is rather straightforward to write down the explicit form of the traces \eqref{gravitytraces} and \eqref{mattertraces}. Introducing the convenient abbreviation
\be
\cV \equiv \frac{d!}{2 \, (4 \pi)^{d/2} \, \Gamma(d/2+1)} \left( \frac{r}{d(d-1)} \right)^{d/2}
\ee
and using that the volume of the $d$-sphere may be written as $V_d = \tfrac{2}{d!} \Gamma(d/2+1) (4 \pi)^{d/2} a^d$ one sees that $V_d \, k^d \, \cV = 1$, which then allows to extract the volume factor from the traces rather easily. The matter traces \eqref{mattertraces} then evaluate to
\begin{subequations}\label{matterevaluated}
	\begin{align}
	T^{\rm scalar} = & \, V_d \, k^d \,  \cV \, \frac{N_S}{1+ \alpha_S^M r} \, T_d^{(0)}(N) \, , \\
	T^{\rm Dirac} = & \, - V_d \, k^d \,  \cV \, \frac{N_D}{1 + \left(\alpha^M_D + \tfrac{1}{4}\right) r} \, T_d^{(1/2)}(N) \, , \\
	T^{\rm vector} = & \, V_d \, k^d \,  \cV \, N_V \left(\frac{1}{1 + (\alpha^M_{V_1} + \tfrac{1}{d})r} \, T^{(1)}_d(N) - \frac{1}{1 + \alpha^M_{V_2}r} \left(T_d^{(0)}(N) - 1 \right) \right) \, . 
	\end{align}
\end{subequations}
Here $N$ represents the cutoff obtained from the corresponding spin representation. The last term in $T^{\rm vector}$ originates from removing the lowest scalar eigenmode from the trace encoding the contributions of the longitudinal vector field. The evaluation of the gravitational traces proceeds along the same lines. Denoting derivatives with respect to the RG time $t$ by a dot, one has
\begin{subequations}\label{gravevaluated}
	\begin{align}
	T^{\rm TT} = & \, \tfrac{1}{2} \, \tfrac{V_d \, k^d \, \cV}{f_k^\prime \left(1 + \left(\alpha^G_T + \tfrac{2}{d(d-1)}\right)r\right)}
	 \left(
	\big( (1+\alpha^G_T r) \dot{f}^\prime_k + 2 f^\prime_k \big) \, T^{(2)}_d(N) 
	- k^{-2} \, \dot{f}_k^\prime \, \widetilde{T}^{(2)}_d(N) 
	\right) 
\, , \\
	T^{\rm ghost} = & \, - \tfrac{V_d \, k^d \, \cV}{1 + \left( \alpha_V^G - \tfrac{1}{d} \right) r} \left(T^{(1)}_d(N) - \tfrac{1}{2}d(d+1) \right) \, , \\
	T^{\rm sinv} = & \, \frac{1}{2} \, \tfrac{V_d \, k^d \, \cV}{f_k^{\prime\prime} \left(1 + \big(\alpha^G_S - \tfrac{1}{d-1}\big)r \right) + \tfrac{d-2}{2(d-1) k^2} \, f_k^\prime}
\left(
\big( (1+\alpha^G_S r) \dot{f}^{\prime\prime}_k + 2 f^{\prime\prime}_k \big) \, T^{(0)}_d(N) 
- k^{-2} \, \dot{f}_k^{\prime\prime} \, \widetilde{T}^{(0)}_d(N) 
\right) \, . 	
	\end{align}
\end{subequations}
Here $T^{\rm sinv}$ contains the contribution from \emph{all scalar modes}. The two lowest eigenmodes are removed by adding
\be
\begin{split}
\Delta 	T^{\rm sinv} = & \,- \frac{1}{2} \, \tfrac{V_d \, k^d \, \cV}{f_k^{\prime\prime} \left(1 + \big(\alpha^G_S - \tfrac{1}{d-1}\big)r \right) + \tfrac{d-2}{2(d-1) k^2} \, f_k^\prime}
\left(
\big( (1+\alpha^G_S r) \dot{f}^{\prime\prime}_k + 2 f^{\prime\prime}_k \big) \, (d+2)
- \tfrac{d+1}{d-1} \, r \, \dot{f}_k^{\prime\prime} 
\right) \,
\end{split}
\ee
to the flow equation. Based on the explicit results for the traces, the flow equation for $f_k(R)$ can be written as
\be\label{pdfgeneral}
\begin{split}
V_d \, \dot{f}_k = T^{\rm TT} + T^{\rm ghost} + T^{\rm sinv} + \Delta 	T^{\rm sinv} + T^{\rm scalar} + T^{\rm Dirac} + T^{\rm vector} \, . 
\end{split}
\ee
Note that this result is valid for general dimension $d$, retains the dependence on all endomorphism parameters and can easily be adapted to any interpolation scheme by specifying the corresponding expressions for $N$ according to \eqref{Ncutoff}. The partial differential equation \eqref{pdfgeneral} constitutes the main result of this section. It generalizes the construction \cite{Ohta:2015efa,Ohta:2015fcu} to general dimension $d$ and the presence of minimally coupled matter fields.

In order to facilitate the further analysis, we also note the explicit from of \eqref{pdfgeneral} in $d=4$ and the averaged interpolation used in \cite{Ohta:2015efa,Ohta:2015fcu}. The result is conveniently written in terms of the dimensionless quantities
\be
r = \Rb k^{-2} \, , \qquad \varphi_k(r) = k^{-d} f_k(\Rb) \, .
\ee
Following the structure \eqref{pdfgeneral} one has
\be\label{pdf4d}
\begin{split}
& \, \dot{\varphi} + 4 \varphi - 2 r \varphi^\prime = \cT^{\rm TT} + \cT^{\rm ghost} + \cT^{\rm sinv} + \cT^{\rm scalar} + \cT^{\rm Dirac} + \cT^{\rm vector} \, . 
\end{split}
\ee
Here the $\cT$ constitute the dimensionless counterparts of the traces $T$ divided by the factor $V_d k^d$ and given by
\begin{subequations}\label{gravflow}
	\begin{align}\label{gravTT}
	\cT^{\rm TT} = & \, 
	\tfrac{5}{2 (4\pi)^2} \, \tfrac{1}{1 + \left(\alpha^G_T + \tfrac{1}{6}\right)r} \left(1 + \left(\alpha^G_T - \tfrac{1}{6}\right)r\right)\left(1 + \left(\alpha^G_T - \tfrac{1}{12}\right)r\right) \\ \nonumber & \; \;
		+ \tfrac{5}{12 (4\pi)^2} \, \tfrac{\dot{\varphi}^\prime + 2 \varphi^\prime - 2 r \varphi^{\prime\prime}}{\varphi^\prime} \left(1 + \left(\alpha^G_T - \tfrac{2}{3}\right)r\right)\left(1 + \left(\alpha^G_T - \tfrac{1}{6}\right)r\right) \, , \\ \label{gravsinv}
	\cT^{\rm sinv} = & \, 
	\tfrac{1}{2 (4\pi)^2} 
	\tfrac{ \varphi^{\prime\prime}}{\left(1+ \left(\alpha^G_S - \tfrac{1}{3}\right)r\right) \varphi^{\prime\prime} + \tfrac{1}{3} \varphi^\prime} 
	\left(1 + \left(\alpha^G_S - \tfrac{1}{2}\right)r\right)
	\left(1 + \left(\alpha^G_s + \tfrac{11}{12}\right)r\right)
	 \\ \nonumber & \; \;
	 + \tfrac{1}{12 (4\pi)^2} 
	 \tfrac{\dot{\varphi}^{\prime\prime} - 2 r \varphi^{\prime\prime\prime}}{\left(1+ \left(\alpha^G_S - \tfrac{1}{3}\right)r\right) \varphi^{\prime\prime} + \tfrac{1}{3} \varphi^\prime} 
	 \left(1 + \left(\alpha^G_S + \tfrac{3}{2}\right)r\right)
	 \left(1 + \left(\alpha^G_s - \tfrac{1}{3}\right)r\right)
	 \left(1 + \left(\alpha^G_S - \tfrac{5}{6}\right)r\right) \, , 
	 \\
	\cT^{\rm ghost} = & \, - \tfrac{1}{48 (4\pi)^2} \, \tfrac{1}{1 + (\alpha^G_V - \tfrac{1}{4})r} \, \left( 72 + 18 r (1 + 8 \alpha^G_V) - r^2 (19 - 18 \alpha^G_V - 72 (\alpha^G_V)^2) \right) \, , 
	\end{align}
\end{subequations}
together with the matter results
\begin{subequations}\label{matterflow}
	\begin{align}
\cT^{\rm scalar} = & \, \frac{N_S}{2 (4\pi)^2} \, \frac{1}{1 + \alpha_S^M r} \,  \left( 1 + \left( \alpha_S^M + \tfrac{1}{4} \right) r \right) \left(1 + \left( \alpha_S^M +  \tfrac{1}{6} \right) r\right) \, , \\ \label{Tdirac}
\cT^{\rm Dirac} = & \, - \frac{2 N_D}{(4\pi)^2} \, \left(1 + \left(\alpha^M_D + \tfrac{1}{6}\right)r\right) \, , \\
\cT^{\rm vector} = & \, \frac{N_V}{2 (4\pi)^2} \, \bigg(
\tfrac{3}{1 + \left( \alpha^M_{V_1} + \tfrac{1}{4} \right) r} 
\left(1 + \left(\alpha^M_{V_1} + \tfrac{1}{6} \right) r \right) 
\left(1 + \left(\alpha^M_{V_1} + \tfrac{1}{12} \right)r\right) \\ & \qquad \qquad \nonumber
- \tfrac{1}{1 + \alpha^M_{V_2} r} \left( 1 + (\alpha^M_{V_2} + \tfrac{1}{2}) r \right) \left(1 + ( \alpha^M_{V_2} - \tfrac{1}{12}) r\right) 
\bigg) \, .  
	\end{align}
\end{subequations}
Here primes and dots denote derivatives with respect to $r$ and $t$, respectively, and all arguments and subscripts have been suppressed in order to aid the readability of the expressions. The result \eqref{gravflow} agrees with the beta functions reported in \cite{Ohta:2015fcu} and \eqref{matterflow} constitutes its natural extension to minimally coupled matter fields. With the result \eqref{pdf4d} at our disposal, we now have all the prerequisites to study the fixed point structure of gravity-matter systems at the level of $f(R)$-gravity. 
 
We close this section by highlighting the central properties of \eqref{pdfgeneral}. Inspecting the gravitational sector \eqref{gravflow} one finds that the function $\varphi_k$ enters the traces in form of its first, second, and third derivative with respect to $r$. As a consequence, a constant term in $\varphi_k$ does not appear on the right-hand-side of the flow equation. This implies in particular that the propagators of the fluctuation fields do not contain contributions from a cosmological constant. This particular feature is owed to the interplay of the exponential split (removing the contribution of the cosmological constant 
from the propagator of the transverse-traceless fluctuations) and the physical gauge $\beta \rightarrow -\infty$ removing the $hh$-term from the gravitational sector \eqref{HessianGrav}.

For the specific regulator \eqref{RLitim}, the evaluation of the spectral sums $S$ results in polynomials that are at most quadratic in $r$ while the sums within $\widetilde{S}$ terminate at order $r^3$. This feature has already been observed in \cite{Codello:2007bd,Machado:2007ea} where it was found that evaluating the flow equation for $f(R)$-gravity for a Litim-type regulator required the knowledge of a finite number of heat-kernel coefficients only. In this sense, it is expected that the Litim regulator leads to similar features when evaluating the operator traces as spectral sums. 

An interesting feature of the averaged interpolation is that the contribution of the Dirac fermions is given by \emph{a polynomial of first order} in the dimensionless curvature $r$. This particular property can be traced to a highly non-trivial cancellation between the propagator and the factors of the spectral sum $T^{(1/2)}_d$. As a consequence, the Dirac fields will not contribute to the flow equation at order $r^2$ and higher. This particular feature is specific to the averaged interpolation and absent in other interpolation schemes (cf.\  \eqref{Tdirac1} and \eqref{Tdirac2} in Appendix \ref{App.A}). Owed to the investigation of the fixed point properties in terms of the matter deformation parameters introduced in \eqref{dgdl} and \eqref{dbetadef} this feature will not be essential when studying non-trivial renormalization group fixed points in the sequel.

Setting the derivatives with respect to the RG time to zero, \eqref{pdf4d} reduces to a third order differential equation for $\varphi_*(r)$. The order of the equation is determined by the scalar contribution arising in the gravitational sector. Casting the resulting expression into normal form by solving for $\varphi^{\prime\prime\prime}$ one finds that that the equation possesses four fixed singularities situated at
\be\label{fixedsing}
r^{\rm sing}_1 = - \frac{1}{\alpha^G_S + \tfrac{3}{2}} \, , \quad
r^{\rm sing}_2 = 0 \, , \quad 
r^{\rm sing}_3 = - \frac{1}{\alpha^G_S -\tfrac{5}{6}} \, , \quad  
r^{\rm sing}_4 = - \frac{1}{\alpha^G_S - \tfrac{1}{3}}  \, . 
\ee
Solutions obtained from solving the differential equation (in normal form) numerically are typically well-defined on the intervals bounded by these singular loci only. Extending a solution across a singularity puts non-trivial conditions on the initial conditions of the fixed point equation. Based on the singularity counting argument \cite{Dietz:2012ic}, stating that each first order pole on the interval $r \in [0,\infty[$ fixes one free parameter, it is then expected that \eqref{pdf4d} admits a discrete set of global fixed functionals. 


\section{Fixed point structure of $f(R)$-gravity matter systems}
\label{sec:fixedpoints}
In this section we discuss the fixed point structure of \eqref{pdf4d} arising within polynomial approximations of the function $\varphi_k(r)$. The general framework is introduced in Sect.\ \ref{sect.41} while the fixed point structure arising at the level of the Einstein-Hilbert truncation and polynomial approximations up to order $N=14$ are investigated in Sects.\ \ref{sect.42} and \ref{sect.43}, respectively. The main focus is on matter sectors containing the field content of the standard model of particle physics (SM) and its most commonly studied phenomenologically motivated extensions (cf.\ Tables \ref{Tab.3} and \ref{Tab.mainresults}).
\subsection{Polynomial $f(R)$-truncation: general framework}
\label{sect.41}
The key ingredient in realizing the asymptotic safety mechanism is a non-Gaussian fixed point (NGFP) of the theories RG flow. At the level of the partial differential equation \eqref{pdf4d}, such fixed points correspond to global, isolated, and $k$ stationary solutions $\varphi_*(r)$. The existence and properties of such fixed functionals will be subject to a companion paper \cite{Alkofer:inprep}. Since the primary focus on the present work is on understanding the predictive power of gravity-matter systems in the gravitational sector, we follow a different strategy and perform an expansion of $\varphi_k(r)$ in powers of $r$, terminating the series at a finite order $r^N$: 
\be\label{polyexpansion}
\varphi_k(r) = \frac{1}{(4\pi)^2} \sum_{n=0}^N \, g_{n}(k) \, r^n \, , \qquad \dot{\varphi}_k(r) = \frac{1}{(4\pi)^2} \sum_{n=0}^N \, \beta_{g_n} \, r^n \, . 
\ee
By construction, the $k$-dependent dimensionless couplings $g_n(k)$, $n=0, \ldots, N$ satisfy
\be
\partial_t g_n \equiv \beta_{g_n} \, , \qquad n=0, \ldots, N \, .
\ee

The explicit expressions for the $\beta_{g_n}$ as a function of the couplings are obtained as follows. First, the ansatz \eqref{polyexpansion} is substituted into  \eqref{pdf4d} which is subsequently expanded in powers of the dimensionless curvature $r$ up to order $r^N$. Equating the coefficients multiplying the terms proportional to $r^n$, $n=0, \ldots N$ results in $N+1$ equations depending on $\beta_{g_n}$ and $g_n$, $n=0, \ldots N$. Solving this system of algebraic equations for $\beta_{g_n}$ determines the beta functions as a function of the couplings $g_n$. Since the resulting algebra is straightforward but quickly turns lengthy, these manipulations are conveniently done by a computer algebra program.

The most important property of the beta functions $\beta_{g_n}(g_0, \ldots, g_N)$ are their fixed points $g^* = \{g_0^*, \ldots, g_N^*\}$ where, by definition,
\be
\left. \beta_{g_n}(g_0, \ldots, g_N)\right|_{g = g^*} = 0 \, , \qquad \forall \, n = 0, \ldots, N \, .
\ee
If an RG trajectory approaches such a fixed point at high energy, the fixed point allows to remove the UV cutoff without introducing divergences in the dimensionless couplings. In this way the theory is rendered asymptotically safe.

The predictive power of an RG fixed point then depends on the dimension of the set of RG trajectories approaching it as $k\rightarrow \infty$. This information is conveniently obtained from the stability matrix
\be\label{Bmat}
{\bf B}_{nm} \equiv \left. \partial_{g_m} \, \beta_{g_n}\right|_{g = g^*}
\ee
which governs the linearized RG flow in the vicinity of the fixed point. Defining the stability coefficients $\theta_n$ as minus the eigenvalues of ${\bf B}$, eigendirections with ${\rm Re}(\theta_n) > 0$ (${\rm Re}(\theta_n) <0$) attract (repel) the flow as $k \rightarrow \infty$. Thus stability coefficients with positive real part are linked to ``relevant directions'' associated with free parameters which have to be determined experimentally. Ideally, fixed points underlying an asymptotic safety construction should come with a low number of free parameters. This implies in particular that the number of relevant directions should saturate when the order of the polynomials appearing in \eqref{polyexpansion} exceeds a certain threshold in $N$. For pure gravity, this test has been implemented in the seminal works \cite{Codello:2007bd,Machado:2007ea} and later on extended in \cite{Codello:2008vh,Falls:2013bv,Falls:2014tra,Falls:2017lst}. A systematic investigation for gravity-matter systems is still missing though. In the remainder of this section, we then follow a two-fold search strategy. In Sect.\ \ref{sect.42} we first identify the matter sectors which give rise to a suitable NGFP at the level of the Einstein-Hilbert action. Based on these initial seeds the stability of these NGFPs under the addition of higher-order scalar curvature terms for phenomenologically interesting gravity-matter systems is investigated in Sect.\ \ref{sect.43}.
 
The fact that the right-hand-side of \eqref{pdf4d} is independent of $g_0$ thereby leads to the 
peculiar feature that $\partial_\lambda \beta_\lambda|_{g = g_*} = -4$ while  $\partial_\lambda \beta_{g_n} = 0$. This structure ensures that the stability matrix always gives rise to a stability coefficient $\theta_0 = 4$, independent of the order $N$ of the polynomial expansion.  
 
\subsection{Surveying the fixed point structure in the Einstein-Hilbert truncation}
\label{sect.42}
We start by analyzing the fixed point structure entailed by \eqref{pdf4d} at the level of the Einstein-Hilbert truncation. In this case the function $\varphi_k(r)$ is approximated by a polynomial of order one in $r$,
\be\label{EHansatz}
\varphi_k(r) = \frac{1}{16 \pi \, g_k} \, \left( 2 \lambda_k - r \right) \, . 
\ee 
The scale-dependent dimensionless cosmological constant $\lambda_k$ and Newton's constant $g_k$ are related to their dimensionful counterparts $\Lambda_k$ and $G_k$ by $\Lambda_k = \lambda_k \, k^2$ and $G_k = g_k \, k^{-2}$. The beta functions controlling the scale-dependence of $g_k$ and $\lambda_k$ in the presence of an arbitrary number of minimally coupled matter fields are readily obtained from substituting the ansatz \eqref{EHansatz} into the partial differential equation \eqref{pdf4d} and projecting the result onto the terms independent of and linear in $r$, respectively. The resulting equations take the form
\be
\partial_t \lambda_k = \beta_\lambda(g_k,\lambda_k) \, , \qquad \partial_t g_k = \beta_g(g_k,\lambda_k) \, , 
\ee
where
\be\label{betaEH}
	\beta_\lambda = - \left(2 - \eta_N \right) \lambda + \frac{g}{24 \pi} \left( 12 - 5 \eta_N + 6 d_\lambda \right) \, , \qquad \beta_g = \left(2+\eta_N\right)g \, . 	
\ee
The anomalous dimension of Newton's constant, $\eta_N \equiv G_k^{-1} \partial_t G_k$ can be cast into the standard form \cite{Reuter:1996cp}
\be
\eta_N = \frac{g \, B_1}{1 - g \, B_2}
\ee
where $B_1$ and $B_2$ are $\lambda$-independent coefficients depending on the choice of coarse graining operator,
\begin{subequations}
	\begin{align}
	\mbox{Type I:} \qquad  & B_1 = - \frac{1}{24 \pi} \left(43 - 4 d_g \right) \, ,  \qquad  B_2 = \frac{25}{72 \pi} \\
	\mbox{Type II:} \qquad & B_1 = - \frac{1}{24 \pi} \left(62 - 4 d_g \right) \, ,  \qquad  B_2 = \frac{35}{72 \pi} \, , 
	\end{align}
\end{subequations}
and the parameters $d_g$ and $d_\lambda$ summarize the matter content of the model
\be\label{dgdl}
d_\lambda =  N_S + 2 N_V -4 N_D \, , \qquad
\begin{array}{ll}
	\mbox{Type I:} \; & d_g = \tfrac{5}{4} N_S - \tfrac{5}{4} N_V - 2 N_D \\[1.1ex]
	\mbox{Type II:} \; & d_g = \tfrac{5}{4} N_S - \tfrac{7}{2} N_V + N_D
\end{array} \, . 
\ee

At this stage the following remarks are in order. The expression for $d_\lambda$ is independent of the choice of coarse graining operator and agrees with the heat-kernel based computations \cite{Codello:2008vh}. Essentially, $d_\lambda$ entails that each bosonic degree of freedom contributes to the running of the cosmological constant with a weight $g/(4\pi)$  while each fermionic degree of freedom contributes with the same factor but opposite sign. The results for $d_g$ differ from the ones based on the early-time expansion of the heat-kernel \cite{Codello:2008vh} where $d_g^{\rm Type \, I} = N_S - N_V - N_D$ and $d_g^{\rm Type \, II} = N_S - 4 N_V + 2 N_D$. This feature just reflects the fact that the evaluation of the spectral sums based on the averaged staircase agrees with the early-time expansion of the heat-kernel at leading order only. One observes, however, that for both choices of coarse-graining operator all fields contribute with their characteristic signature, so that the resulting picture may be qualitatively similar.

As a second remarkable feature, the beta functions \eqref{betaEH} do not contain denominators of the form $(1-c \lambda)^n$ ($c > 0$) which typically lead to a termination of the flow at a finite value $\lambda = 1/c$ \cite{Reuter:2001ag}. As a consequence the flow is well-defined for any value of $\lambda$ and gives rise to a globally well-defined flow diagram \cite{Gies:2015tca}. Moreover, the mechanism for gravitational catalysis \cite{Wetterich:2017ixo} is not realized in the present framework.

Owed to their simple algebraic structure, the fixed points of the beta functions \eqref{betaEH} can be found analytically. They possess a Gaussian fixed point (GFP) located at $(g^*,\lambda^*) = (0,0)$ whose stability coefficients are given by the canonical mass dimension of the dimensionful Newton's constant and cosmological constant. In addition the system exhibits \emph{a single NGFP for any given matter sector}. For a coarse graining operator of Type I (vanishing endomorphisms) this fixed point is situated at
\be\label{EH:NGFP}
g^* = \frac{144 \pi}{179 - 12 \, d_g} \, , \qquad \lambda^* = \frac{33+9 d_\lambda}{179-12 \, d_g} \, ,
\ee
while its stability coefficients obtained from \eqref{Bmat} are
\be\label{EH:theta}
\theta_0 = 4 \, , \qquad \theta_1 = \frac{358 - 24 \, d_g}{3 \, (43-4 \, d_g)} \, . 
\ee
The corresponding expressions for the Type II coarse graining operator are obtained along the same lines and have a similar structure. 

The properties of the NGFPs \eqref{EH:NGFP} as a function of the matter content are illustrated in Fig.\ \ref{fig.3}.
\begin{figure}[t!]
	\includegraphics[width=0.48\textwidth]{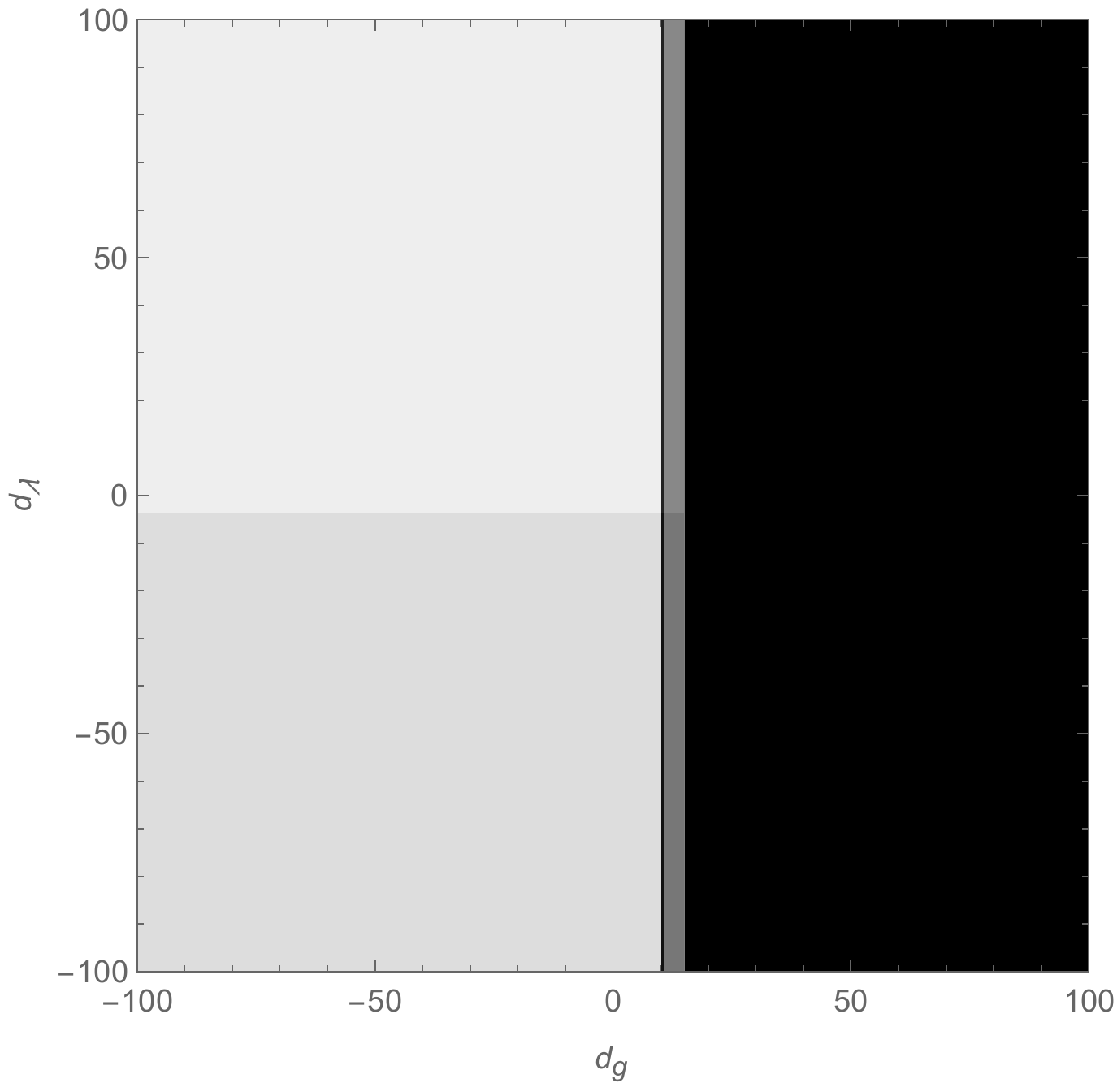}
	\includegraphics[width=0.48\textwidth]{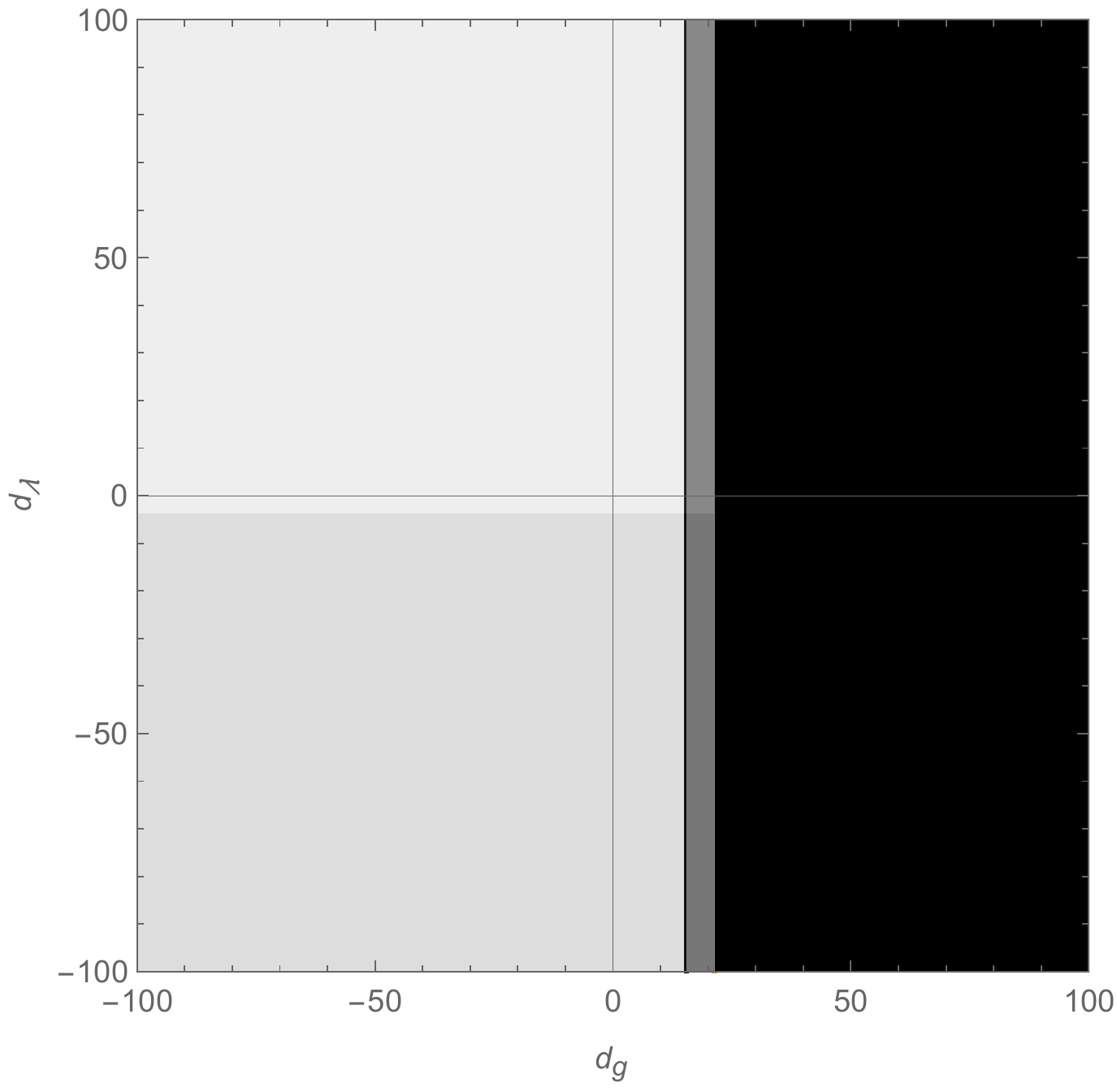}
	\caption{Illustration of the fixed point structure arising from the system \eqref{pdf4d} at the level of the Einstein-Hilbert truncation for a coarse graining operator of Type I (left) and Type II (right) respectively. The matter content of the model is encoded in the parameters $d_g, d_\lambda$ defined in  \eqref{dgdl}. The black region does not support a NGFP with $g_* > 0$. In the dark gray region the NGFP is a saddle point with $\theta_1 < 0$. The gray and light gray regions support a UV attractive NGFP with $g_* \lambda_* < 0$ and $g_* \lambda_* > 0$, respectively. Notably, the qualitative fixed point structure, classified in terms of $d_g, d_\lambda$ is independent of the choice of coarse graining operator. \label{fig.3}}
\end{figure}
Quite remarkably, the fixed point structure resulting from the Type I and Type II coarse graining operator is \emph{qualitatively identical} provided that the matter content of the model is encoded in the deformation parameters \eqref{dgdl}.
It is determined by three separation lines, $L_1, L_2, L_3$ situated at
\be
	\begin{tabular}{llll}	\mbox{Type I:}  & \qquad $L_1: \, d_g = \tfrac{179}{12}$\, ,  & \qquad $L_2: \, d_g = \tfrac{43}{4}$\, ,  & \qquad $L_3: \, d_\lambda = - \tfrac{11}{3}$ \, ,  \\
	\mbox{Type II:} &  \qquad $L_1: \, d_g = \tfrac{64}{3}$\, ,  & \qquad $L_2: \, d_g = \tfrac{31}{2}$\, , &  \qquad $L_3: \, d_\lambda = - \tfrac{11}{3}$ \, . 
	\end{tabular}
\ee
For matter sectors located to the left (right) of $L_1$ the NGFP is situated at $g^* > 0$ ($g^* < 0$), respectively. If $g^* < 0$, the corresponding fixed point is disconnected from the physically viable low-energy regime and may therefore not be suitable for controlling the high-energy behavior of physically interesting theories. Thus this case will be discarded from the further analysis. Matter sectors sitting in the region bounded by the lines $L_1$ to the right and $L_2$ to the left support a saddle point where $\theta_1 < 0$ while for matter systems to the left of $L_2$ the NGFP is UV-attractive in both $g$ and $\lambda$. The horizontal line $L_3$ separates the regions where the NGFPs come with $\lambda_* g_* < 0$ (lower-left region) and $\lambda_* g_* > 0$ (upper-left region), respectively.

Fig.\ \ref{fig.3} makes it also apparent that the systems where scalar matter is coupled to gravity possesses an upper bound on the number of scalar fields ($N_S^{\rm max} = 14$ for Type I and $N_S^{\rm max} = 21$ for Type II). If $N_S$ exceeds these bounds the NGFP is located in the $g^* < 0$-region. While the fixed point is still present, it is no longer suitable for realizing a phenomenologically interesting gravity-matter system. We will further elaborate on this point in the conclusion section. 

We close the present discussion by summarizing the details for the NGFPs
found for distinguished gravity-matter systems in Table \ref{Tab.3}.
\begin{table}[t!]
	\renewcommand{\arraystretch}{1.4}
	\begin{center}
		\begin{tabular}{p{3.94cm}||c|c|c||c|c|c|c||c|c|c|c} 
		model &	\multicolumn{3}{c||}{matter content} & \multicolumn{4}{c||}{Type I coarse graining} & \multicolumn{4}{c}{Type II coarse graining} \\ 
			& \, $N_S$ \, & \, $N_D$ \, & \, $N_V$ \, & \, $d_g$ \, & \, $d_\lambda$ \, & \, $g_*\lambda_*$ \, &  \, $\theta_1$ \, & \, $d_g$ \, & \, $d_\lambda$ \, & \, $g_*\lambda_*$  \, &  \, $\theta_1$ \, \\  \hline \hline
			pure gravity & 0 & 0 & 0 &  0 & 0 & $0.47$ & $2.78$ & $0$ & $0$ & $0.23$ & $2.75$ \\ \hline
			Standard Model (SM) & 4 & $\tfrac{45}{2}$ & 12 & $-55$ & $-\,62$ & $-0.34$ & $2.13$ & $-\tfrac{29}{2}$ & $-62$ & $-1.28$ & $2.39$  \\ \hline
			SM, dark matter (dm) & 5 & $\tfrac{45}{2}$ & 12 &  $-\,\tfrac{215}{4}$ & $-61$ & $-0.34$ & $2.13$ & $-\tfrac{53}{4}$ & $-61$ & $-1.36$ & $2.41$ \\ \hline
			SM, $3\,\nu$ & 4 & 24 & 12 & $-58$ & $-68$ & $-0.34$ & $2.12$ & $-13$ & $-68$ & $-1.54$ & $2.41$ \\ \hline
			SM, $3\,\nu$, dm, axion & 6 & 24 & 12 & $-\tfrac{111}{2}$ & $-66$ & $-0.36$ & $2.13$ & $- \tfrac{21}{2}$ & $-66$ & $-1.74$ & $2.45$ \\ \hline
			MSSM & 49 & $\tfrac{61}{2}$ & 12 & $-\tfrac{59}{4}$ & $-49$ & $-1.45$ & $2.33$ & $\tfrac{199}{4}$ & $-49$ & $-$ & $-$ \\ \hline
			{SU(5) GUT} & {124} & {24} & {24} &  $77$ & $76$ & $-$ & $-$ & $95$ & $76$ & $-$ & $-$ \\ \hline
			{SO(10) GUT} & {97} & {24} & {45} & $17$ & $91$ & $-$ & $-$ & $-\tfrac{49}{4}$ & $91$ & $2.37$ & $2.42$ \\ \hline \hline
		\end{tabular}
	\end{center}
	\caption{\label{Tab.3} Fixed point structure arising from the field content of commonly studied matter models. The SM and its extensions by a small number of additional matter fields support NGFPs with very similar properties.}
\end{table}
The list covers the cases of pure gravity, gravity coupled to the field content of the standard model of particle physics (SM), and phenomenologically motivated matter sectors arising in frequently studied candidates for physics beyond the standard model. The latter supplement the field content of the SM by additional scalar fields (dark matter (dm) or axion candidates), right-handed neutrinos, supersymmetric partners of the SM fields leading to the minimally supersymmetric standard model (MSSM), or fields required in the realization of grand unified theories (GUTs) based on the gauge groups SU($5$) or SO($10$).  By substituting the matter field content listed in the second to fourth column of Table \ref{Tab.3} into the maps \eqref{dgdl} and checking the resulting coordinates in Fig.\ \ref{fig.3} readily shows that many of these models give rise to a NGFP which is UV attractive for both Newton's constant and the cosmological constant (i.e., $\theta_1 > 0$). The exceptions are the GUT-type models (Type I coarse graining operator) and the MSSM and SU($5$) GUT (Type II coarse graining operator) which lead to NGFPs with $g^* < 0$ and thus fail the test of asymptotic safety at the level of the Einstein-Hilbert truncation. Table \ref{Tab.3} provides the starting ground for investigating which of the gravity-matter fixed points are actually stable when higher-order scalar curvature terms are included in the ansatz for $\varphi_k(r)$. 

\subsection{Gravity-matter fixed points in the presence of an $r^2$-term}
\label{sect.44}
Owed to the special property that the beta functions for the dimensionless couplings $g_n$, $n \ge 1$ are independent of $g_0$, the polynomial expansion \eqref{polyexpansion} to order $N=2$ also gives rise to a two-dimensional subsystem of beta-functions which closes on its own. One may then study the fixed point structure for $g_1$ and $g_2$ arising from 
\be\label{betar2}
\beta_{g_1}(g_1,g_2)|_{g = g^*} = 0 \, , \qquad \beta_{g_2}(g_1,g_2)|_{g = g^*} = 0 \, , 
\ee
for arbitrary matter sectors. Once a fixed point $(g_1^*,g_2^*)$ is obtained its coordinates may be substituted into the beta function $\beta_{g_0}(g_0,g_1,g_2)$. Solving $\beta_{g_0}(g_0^*,g_1^*,g_2^*) = 0$ for $g_0^*$ then determines the value of $g_0$ uniquely.

The triangular shape of the stability matrix furthermore guarantees that the stability coefficients $\theta_1$, $\theta_2$ obtained from the $g_1$-$g_2$ subsystem carry over to the full system. As a result the stability coefficients from the $N=2$ expansion are $\theta_0 = 4, \theta_1, \theta_2$ where the latter depend on the specific matter content and choice of coarse graining operator. 

Following the strategy of the last subsection, we encode the matter contribution to the beta functions \eqref{betar2} by $d_g$, introduced in \eqref{dgdl}, supplemented by
\be\label{dbetadef}
d_\beta \equiv N_S + 2 N_V \, . 
\ee
Note that this parameter is actually independent of the choice of coarse graining operator. Moreover, it is independent of the number of Dirac fields which is owed to the cancellation between numerator and denominator observed in  \eqref{Tdirac}. Since all matter fields contribute to $d_\beta$ with a positive sign all matter models are located in the upper half-plane $d_\beta \ge 0$ with $d_\beta = 0$ realized by pure gravity and gravity coupled to an arbitrary number of Dirac fields. The map $(N_S,N_V,N_D) \mapsto (d_\lambda, d_g, d_\beta)$ is actually bijective such that any particular matter sector is uniquely characterized by either its field content or its coordinates $(d_\lambda, d_g, d_\beta)$. 

Keeping the values of $d_g, d_\beta$ general, the analysis of \eqref{betar2} shows that the reduced system can have at most three (five) solutions for a coarse graining operator of Type I (Type II). Applying the selection criteria that the fixed point coordinates are real and obey $g_1^* < 0$, the number of candidate NGFPs is shown in the left column of Fig.\ \ref{Fig.r2class}.
\begin{figure}[t!]
\centering
\includegraphics[width=0.48\textwidth]{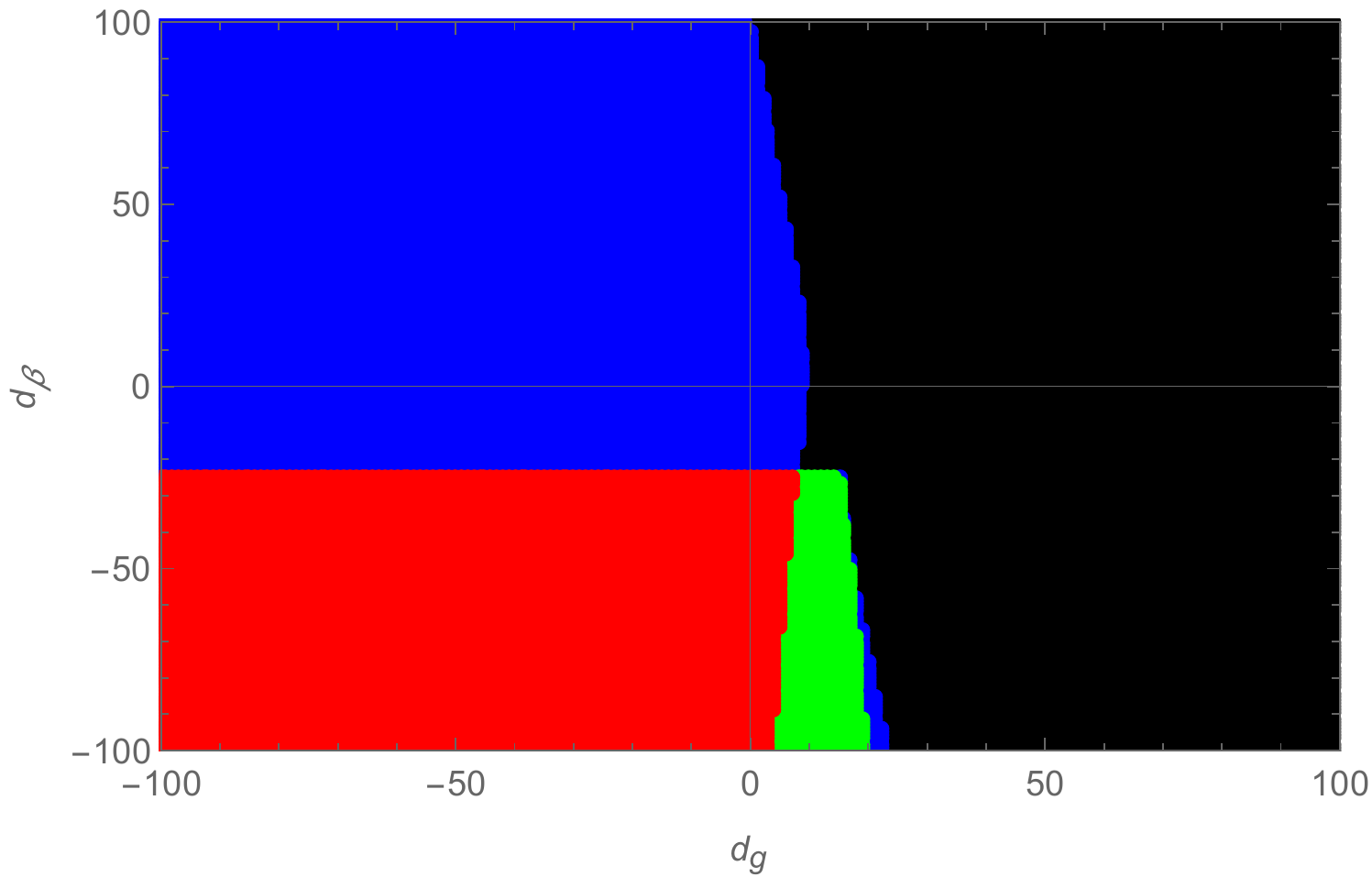} \,
\includegraphics[width=0.48\textwidth]{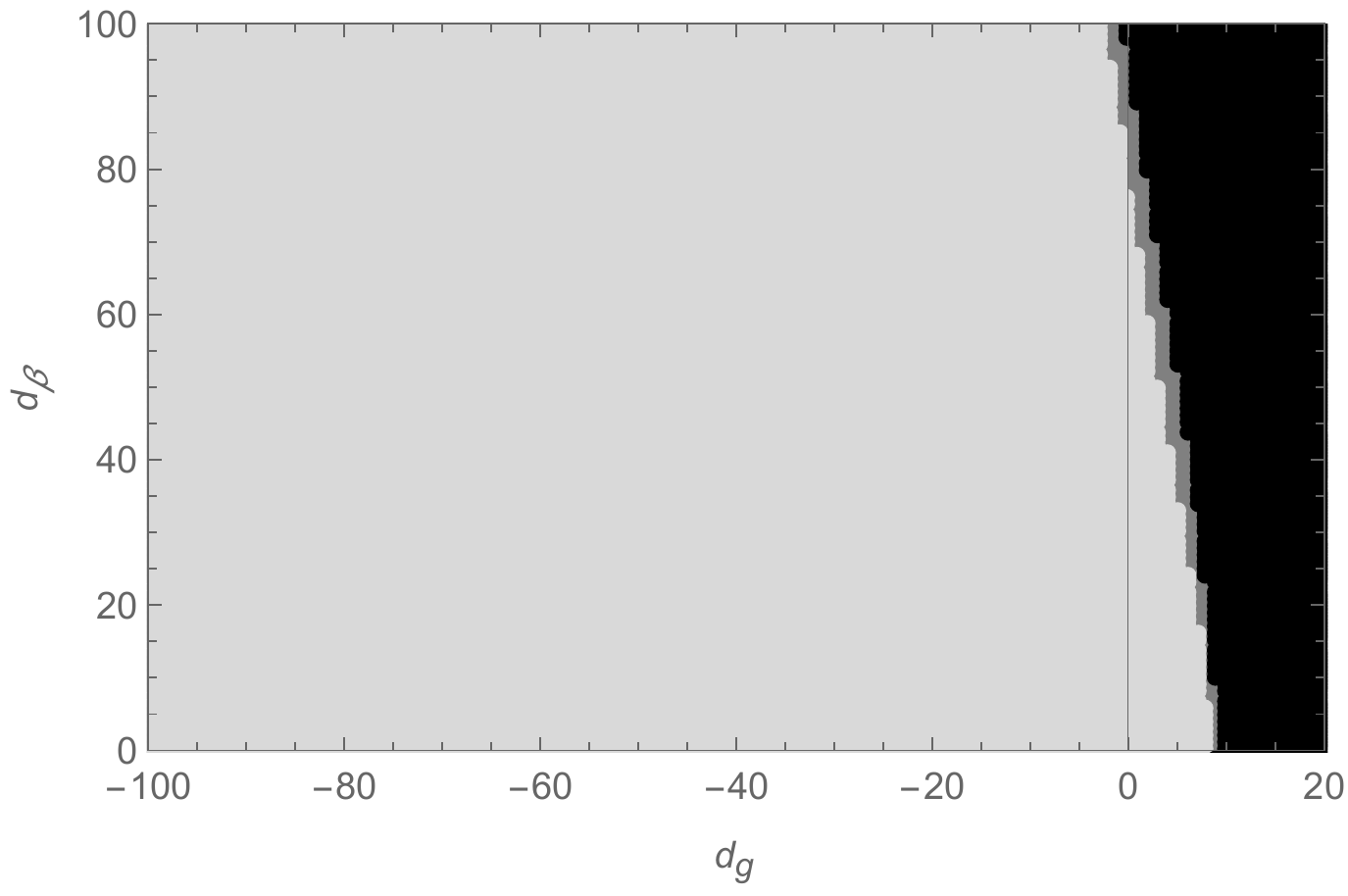} \\
\includegraphics[width=0.48\textwidth]{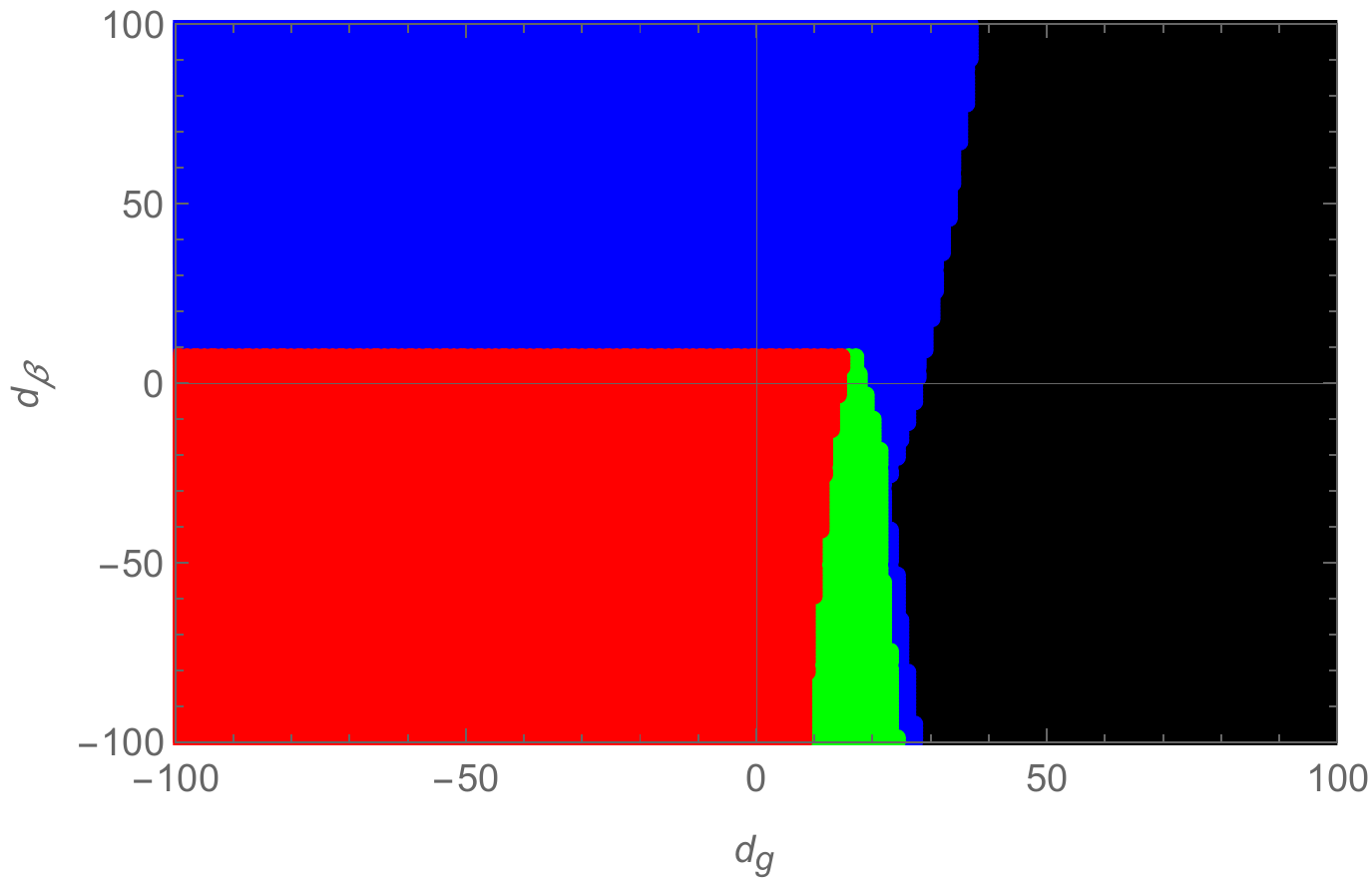} \,
\includegraphics[width=0.48\textwidth]{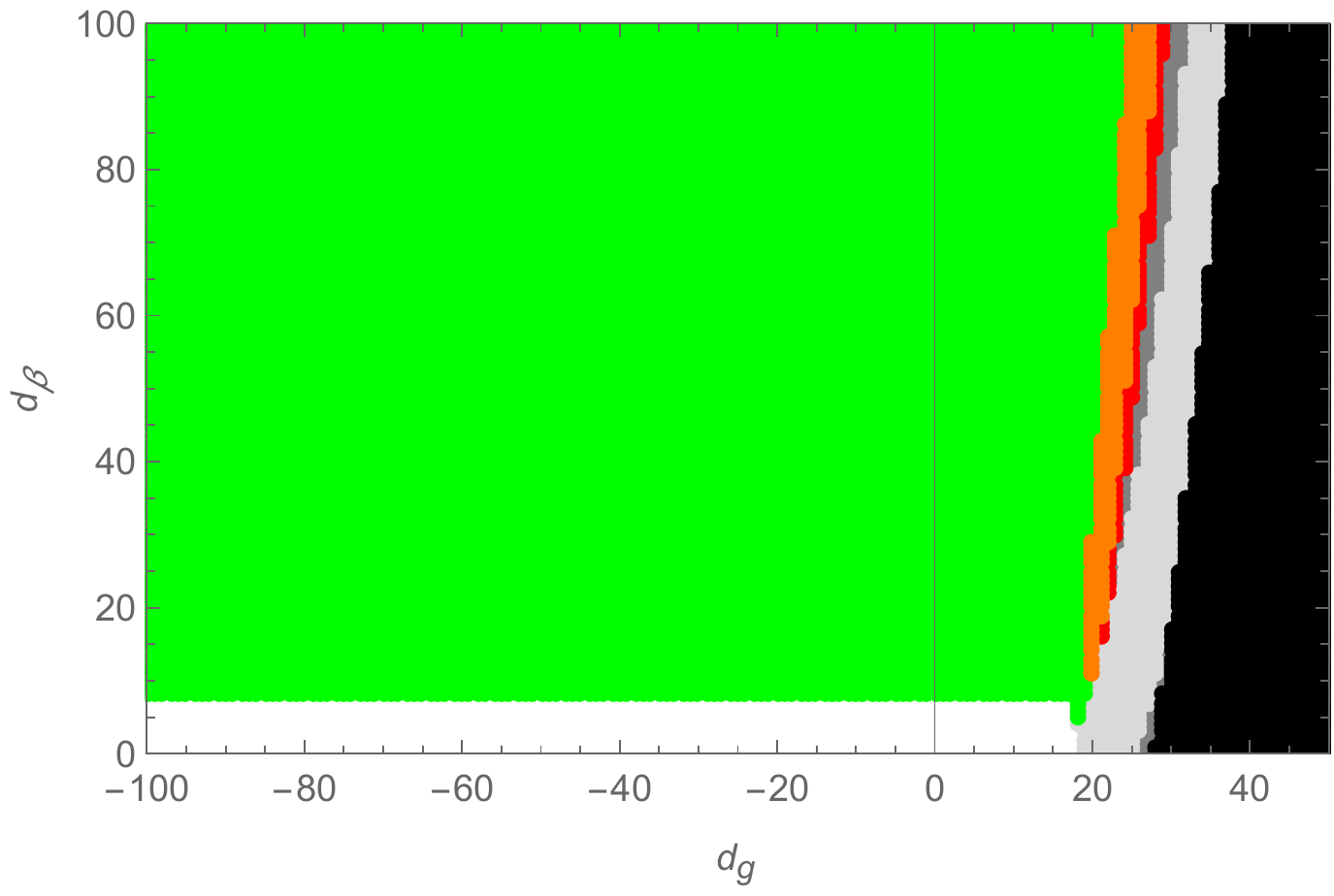}
\caption{\label{Fig.r2class} NGFPs arising in the polynomial $\varphi_k(r)$ approximation at order $N=2$ for a coarse graining operator of Type I (top line) and Type II (bottom line). In the left column the colors black, blue, green, and red indicate that the matter sector supports zero, one, two, and three NGFPs situated in the region with positive Newton's constant. The right column displays the stability properties of the NGFPs with $d_\beta > 0$. For points shaded dark gray, light gray, and green the $\theta_1, \theta_2$ subsystem has zero, one, and two UV-attractive eigendirection with real stability coefficients. In the orange region the eigenvalues of the NGFP are complex.}
\end{figure}
Besides the physically interesting region where $d_\beta \ge 0$, the diagrams also show the fixed point structure for $d_\beta < 0$. In this way it becomes apparent that again, the Type I and Type II coarse graining operator leads to qualitatively similar results. The numerical analysis reveals that there are at most 3 candidate solutions satisfying the selection criteria of a real positive Newton's constant. 

The stability properties of the NGFPs arising from matter sectors supporting a single candidate NGFP are displayed in the right column of Fig.\ \ref{Fig.r2class}. Disregarding the boundary region adjacent to the black region where no admissible NGFP is found reveals an intricate difference between the two coarse graining operators. Focusing on a generic fixed point in the upper-left region one has $\theta_2 < 0$ for Type I and $\theta_2 > 0$ for Type II, i.e., the two cases lead to two and three UV-relevant directions, respectively. The role of the small band of $d_g-d_\beta$-values supporting multiple NGFPs (white region in the lower-right diagram) will be clarified below.

For the matter sectors highlighted in Table \ref{Tab.3}, the addition of the $r^2$-term does not lead to new bounds on the admissible fixed point structure, i.e., all models passing the Einstein-Hilbert test are situated in the region in the $d_g, d_\beta$-plane which supports a unique extension of the fixed point seen for $N=1$ to $N=2$. The sign of the new stability coefficient depends on the choice of coarse graining operator though: in the Type I case $\theta_2 < 0$ while the Type II has $\theta_2 > \theta_1 > 0$ indicating that the new direction is UV relevant with a large, positive stability coefficient.

\begin{figure}[t!]
	\includegraphics[width=0.45\textwidth]{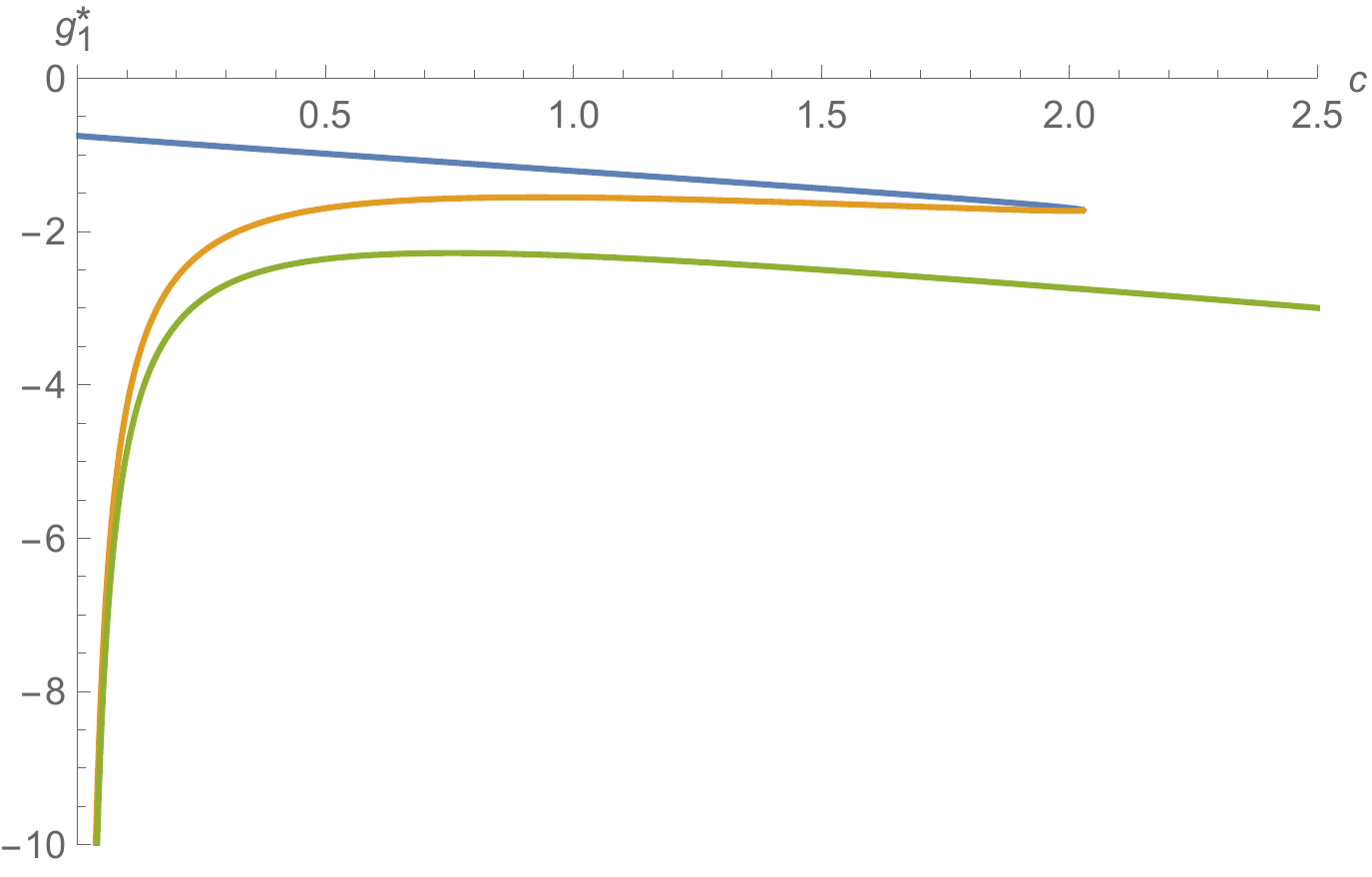} \; \; \;
	\includegraphics[width=0.45\textwidth]{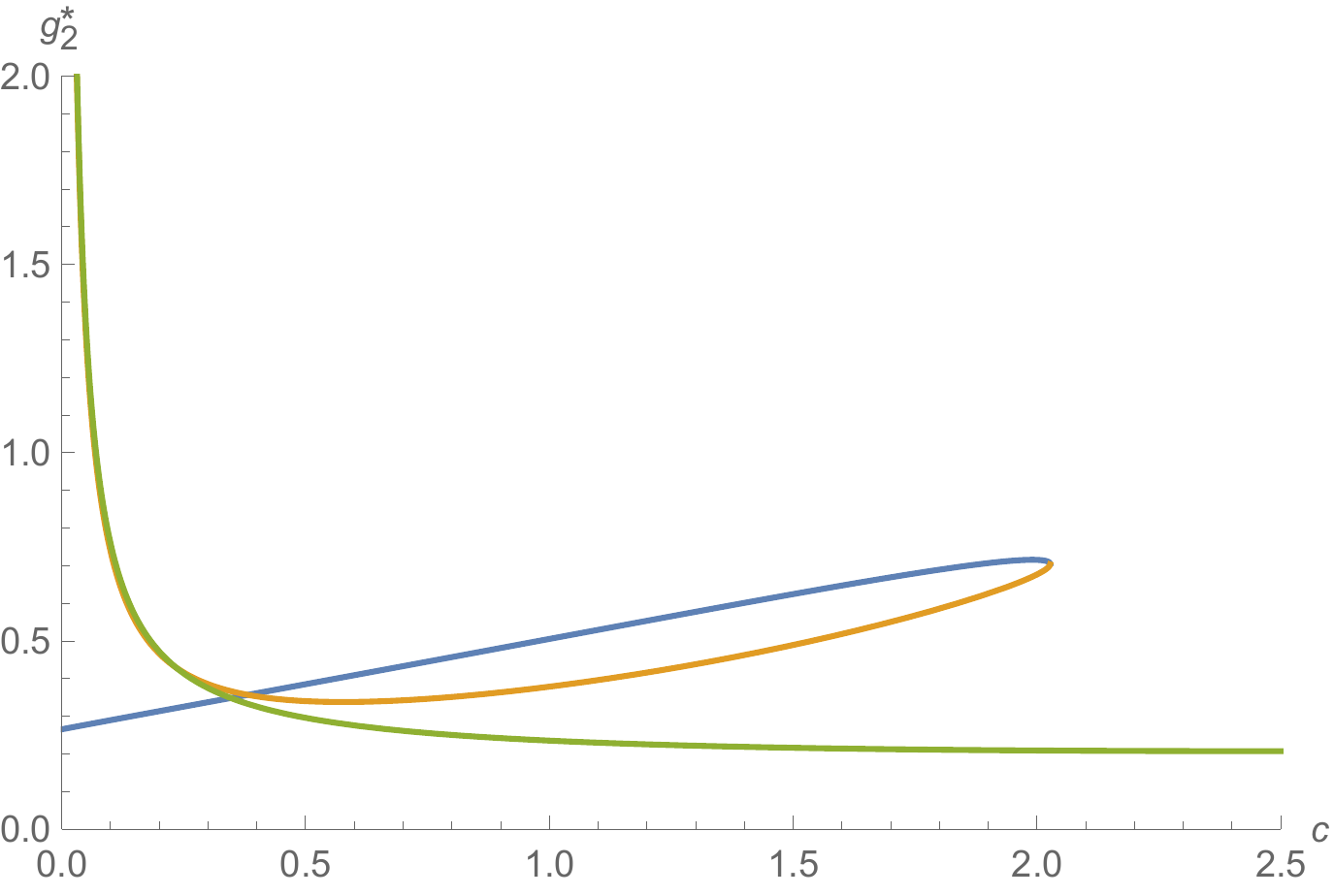} \\[1.4ex]
	\includegraphics[width=0.45\textwidth]{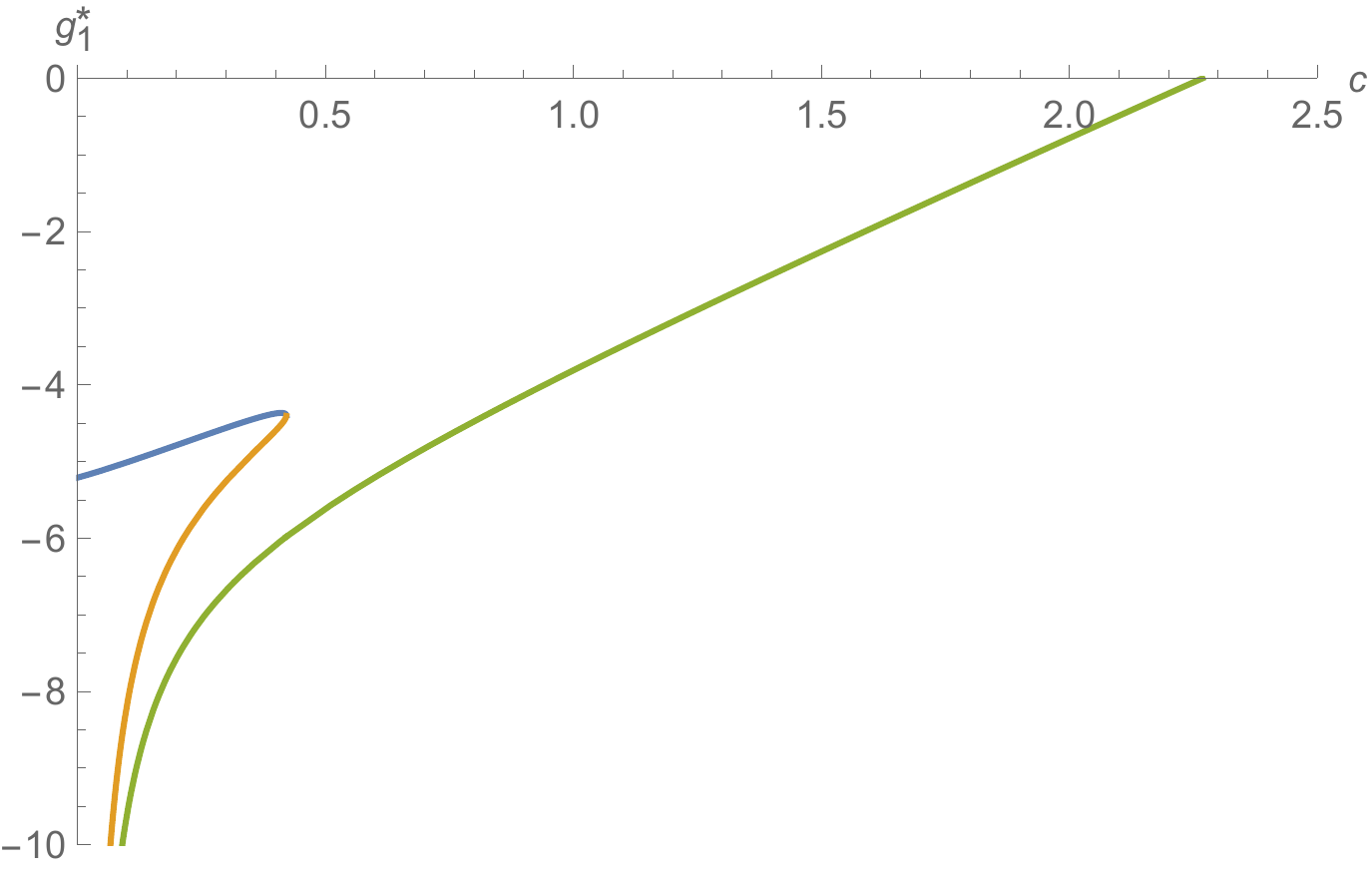}  \; \; \;
	\includegraphics[width=0.45\textwidth]{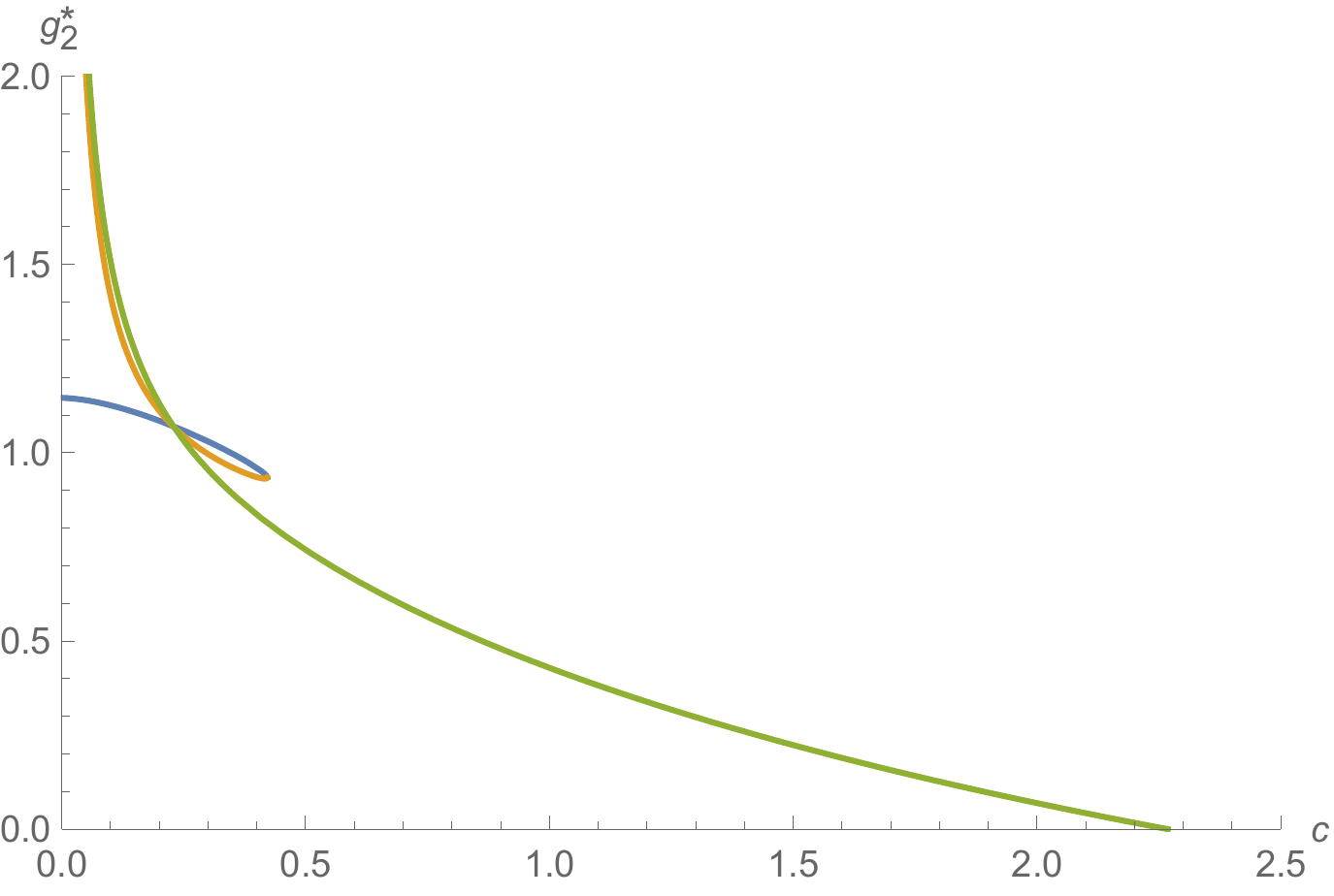}
	\caption{\label{Fig.typeIIannihilation} Fixed point structure obtained for $N=2$ as a function of the deformation parameter $c$ interpolating between a Type I ($c=0$) and Type II ($c=1$) coarse graining. The cases of pure gravity $(N_S = 0, N_D = 0, N_V = 0)$ and the standard model $(N_S = 4, N_D = \tfrac{45}{2}, N_V = 12)$ are shown in the top and bottom row, respectively. The deformation of the NGFP appearing in the Type I analysis is depicted by the blue line. For $c > 0$ there are two additional NGFP moving in from infinite. One of these fixed points annihilates the Type I fixed point at a finite value of $c$. For pure gravity this annihilation occurs at $c > 1$ while for the other gravity-matter models listed in Table \ref{Tab.3} the annihilation is at $c < 0$. As a result the systems resulting from the Type II coarse graining again possess a unique NGFP (green line). This fixed point does not admit a convergent extension to higher orders of $N$, however.}
\end{figure}
At this stage, it is natural to inquire about the relation of the NGFPs seen in the Type I and Type II case. For this purpose, we resort to the interpolating coarse graining operators constructed from \eqref{regtypei}. For $N=2$, the subsystem of equations determining the position of the NGFPs in the $g_1-g_2$-plane is sufficiently simple that all of its five roots can be found for general deformation parameter $c$. The corresponding implicit expressions allow to trace the position of the NGFP seen for Type I coarse graining ($c=0$) as a function of the deformation parameter $c$. Fig.\ \ref{Fig.typeIIannihilation} depicts the $c$-dependence of the fixed point structure obtained for two characteristic examples, pure gravity ($d_\beta = 0$) in the top row and gravity coupled to the matter content of the standard model ($d_\beta = 28$) in the bottom row, respectively. The key structure encountered in the analysis is rather universal. For $c=0$ the system has a single NGFP which is the one displayed in the top line of Fig.\ \ref{Fig.r2class}. Once $c$ is increased an additional pair of NGFPs moves in from infinity (orange and green lines). At a finite value of $c$ one of these new fixed points (orange line) annihilates the $c=0$ solution (blue line). For $c$ larger than this critical value one is again left with a single NGFP (green line). 

If $d_\beta \le 7$ this annihilation occurs at $c > 1$ while for $d_\beta \ge 8$ the two fixed points annihilate before the Type II coarse graining operator is reached. Since all phenomenologically interesting matter sectors are located at $d_\beta \ge 8$ we see that the NGFPs found in the Type II computation \emph{are not continuously connected} to their Type I counterparts. Anticipating results from the next subsection, we will call these two disconnected families of NGFPs to be of ``gravity-type'' (blue line) and ``matter-dominated'' (green line), respectively.  

\subsection{Gravity-matter fixed points for selected matter sectors}
\label{sect.43}
The final part of our analysis investigates the stability of the NGFPs characterized in the previous subsections under the inclusion of further powers of the dimensionless curvature $r$ in the polynomial ansatz \eqref{polyexpansion}. A detailed numerical analysis determining the polynomial solution approximating the fixed point up to $N=14$ and its  critical exponents up to $N=9$ 
revealed a strikingly simple structure: \emph{for gravity-type NGFPs the position and stability coefficients characterizing the fixed point converge rapidly when $N$ is increased. For the matter-dominated NGFPs no such convergence pattern could be established.} In order to arrive at this result extending the order of the polynomials beyond $N=2$ is crucial.

\begin{table}[t!]
	\renewcommand{\arraystretch}{1.4}
	\begin{center}
		\begin{tabular}{p{3.94cm}|c|c|c||c|c||c|c} 
			model &	\multicolumn{3}{c||}{matter content} & \multicolumn{2}{c||}{\; Type I coarse graining \; } & \multicolumn{2}{c}{\; Type II coarse graining \;} \\ 
			& \, $N_S$ \, & \, $N_D$ \, & \, $N_V$ \, & \; \; EH \; \; & \; \; $f(R)$  \; \; &  \; \; EH  \; \; &  \; \; $f(R)$  \; \; \\  \hline \hline
			pure gravity & 0 & 0 & 0 &  $\checkmark$ & $\checkmark$ & $\checkmark$ & $\checkmark$   \\ \hline
			Standard Model (SM) & 4 & $\tfrac{45}{2}$ & 12 & $\checkmark$ & $\checkmark$ & $\checkmark$ & $\left(\xmark\right)$  \\ \hline
			SM, dark matter (dm) & 5 & $\tfrac{45}{2}$ & 12 &  $\checkmark$ &  $\checkmark$ & $\checkmark$ & $\left(\xmark\right)$ \\ \hline
			SM, $3\,\nu$ & 4 & 24 & 12 & $\checkmark$ &  $\checkmark$ & $\checkmark$ & $\left(\xmark\right)$ \\ \hline
			SM, $3\,\nu$, dm, axion & 6 & 24 & 12 & $\checkmark$ &  $\checkmark$ & $\checkmark$ & $\left(\xmark\right)$ \\ \hline
			MSSM & 49 & $\tfrac{61}{2}$ & 12 & $\checkmark$ & $\checkmark$ & $\xmark$ & $\xmark$  \\ \hline
			{SU(5) GUT} & {124} & {24} & {24}  & $\xmark$ & $\xmark$ & $\xmark$ & $\xmark$ \\ \hline
			{SO(10) GUT} & {97} & {24} & {45} & $\xmark$ & $\xmark$ &  $\checkmark$ &  $\left(\xmark\right)$ \\ \hline \hline
		\end{tabular}
	\end{center}
	\caption{\label{Tab.mainresults} Summary of results on the stability of NGFPs appearing for the matter content of the standard model of particle physics and its phenomenologically motivated extensions. Checkmarks $\checkmark$ indicate that the setup possesses a suitable NGFP which converges for increasing $N$. The symbol $\xmark$ shows that there is no NGFP at the level of the Einstein-Hilbert ($N=1$) approximation while a $(\xmark)$ implies that the NGFP seen at $N=1$ does not exhibit convergence when $N$ is increased.}
\end{table}
Table \ref{Tab.mainresults} summarizes the consequences of this general result for phenomenological interesting gravity-matter models introduced in Table \ref{Tab.3}. The key insights are summarized as follows: For pure gravity where a gravity-type NGFP persists for both coarse graining operators, one consequently has one stable NGFP solution in both cases. The characteristics of these NGFPs, including their position and stability coefficients, are tabulated in Tables \ref{Tab.grav.I} and \ref{Tab.grav.II} of Appendix \ref{App.B}, respectively. Focusing on the case of Type I coarse graining and the  gravity-matter models selected in Table \ref{Tab.3}, it is found that all NGFPs seen at the level of the Einstein-Hilbert approximation have a stable extension to polynomial $f(R)$-gravity. For gravity supplemented by the matter content of the standard model, this is strikingly demonstrated in Table \ref{Tab.FPstandardmodel}. 
\begin{table}[p!]
	\centering
	\begin{tabular}{c|ccc|cccc}
		\;\; $N$\;\; & \;  $g_0^*$   \; & \; $g_1^*$ \;  & \; $g_2^*$ \; & \;  $g_3^* \times 10^{-4}$ \; 
		& \; $g_4^* \times 10^{-4}$ \; & \; $g_5^* \times 10^{-4}$ \; & \; $g_6^* \times 10^{-4}$ \; \\ 
		\hline \hline
		$1$ & \; $-7.2917$ \; & \; $-5.8264$ \; & \\
		$2$ & $-6.7744$  & $-5.2122$ & \; $1.1455$ \; \\
		$3$ & $-6.7795$  & $-5.2617$ & $1.1601$ & $50.466$  \\
		$4$ & $-6.7737$  & $-5.2577$ & $1.1550$ & $49.161$ & $-2.7013$ \\
		$5$ & $-6.7742$  & $-5.2598$ & $1.1559$ & $51.122$ & $-2.4926$ & $0.3313$ \\
		$6$ & $-6.7755$  & $-5.2611$ & $1.1571$ & $51.929$ & $-1.9180$ & $0.4268$ & $0.1426$ \\ 
		$7$ & $-6.7764$  & $-5.2632$ & $1.1582$ & $53.712$ & $-1.5336$ & $0.7152$ & $0.1999$ \\ 
		$8$ & $-6.7775$  & $-5.2646$ & $1.1592$ & $54.700$ & $-1.0696$ & $0.8557$ & $0.3065$ \\
		$9$ & $-6.7781$  & $-5.2657$ & $1.1599$ & $55.663$ & $-0.7932$ & $1.0079$ & $0.3586$ \\
		$10$ & $-6.7786$  & $-5.2665$ & $1.1605$ & $56.249$ & $-0.5615$ & $1.0959$ & $0.4091$ \\
		$11$ & $-6.7789$  & $-5.2671$ & $1.1608$ & $56.693$ & $-0.4174$ & $1.1654$ & $0.4382$ \\
		$12$ & $-6.7792$  & $-5.2674$ & $1.1611$ & $56.973$ & $-0.3142$ & $1.2084$ & $0.4602$ \\
		$13$ & $-6.7793$  & $-5.2677$ & $1.1612$ & $56.717$ & $-0.2486$ & $1.2383$ & $0.4737$ \\
		$14$ & $-6.7794$  & $-5.2678$ & $1.1613$ & $55.729$ & $-0.2049$ & $1.2572$ & $0.4830$ \\
		\hline \hline
			\multicolumn{8}{c}{}\\[-2ex]
	\end{tabular}
	
	\begin{tabular}{c|cc|ccccc}
		%
		\; \; $N$\;\; & $\theta_0$ & $\theta_1$ & $\theta_2$ & $\theta_3$ & $\theta_4$ & $\theta_5$ & $\theta_6$ \\ \hline \hline
		$1$ & \;\;\; \;\;\; $4$ \;\;\; \;\;\; & \;\;\; $2.127$ \;\;\;\\
		$2$ & $4$ & $2.339$ & $-1.671$  \\
		$3$ & $4$ & \; $2.274$ \; & $-1.727$ & $-6.013$ \\
		$4$ & $4$ & $2.279$ & \; $-1.808$ \; & \; $-5.905$ \; & \; $-9.308$ \; \\
		$5$ & $4$ & $2.280$ & $-1.809$ & $-5.928$ & $-9.330$ & $-11.956$ 
		\\
		$6$ & $4$ & $2.279$ & $-1.797$ & $-5.916$ &  $-9.297$ & \;  $-12.146$ \;  & \; $-14.293$ \; \\ 
		$7$ & $4$ & $2.278$ & $-1.791$ & $-5.888$ &  $-9.283$ & $-12.070$ & $-14.628$ \\
		$8$ & $4$ & $2.277$ & $-1.784$ & $-5.874$ &  $-9.248$ & $-12.061$ & $-14.519$ \\
		$9$ & $4$ & $2.276$ & $-1.780$ & $-5.856$ &  $-9.225$ & $-12.018$ & $-14.512$ \\
		\hline \hline
	\end{tabular}
	\caption{\label{Tab.FPstandardmodel} Fixed point structure of $f(R)$ gravity coupled to the matter content of the standard model of particle physics $(N_S=4, N_D=45/2, N_V = 12)$ and a Type I cutoff. The fixed point exhibits the same stability properties as in the case of pure gravity. Note that the polynomial coefficients for the constant, linear and quadratic term are of 
		${\mathcal{O}}(1)$ whereas $g_3^*$ is already smaller than 1\% .}
\end{table}
Besides the rapid convergence of the polynomial expansion with regards to the position and stability coefficients of the NGFP, the data shows that the fixed point has the same predictive power as the one found in the case of pure gravity: it comes with two relevant parameters. These characteristic properties are shared by the fixed points found for the other matter sectors carrying ticks in Table \ref{Tab.mainresults}. Their characteristic properties are compiled in Appendix \ref{App.B}.

We close our discussion by highlighting one particular property of the polynomial solutions $\varphi(r)$ associated with the gravity-type NGFPs. Based on the partial differential equation \eqref{pdf4d} one finds that the coefficients $g_0^*$, $g_1^*$ and $g_2^*$ are of order unity with $g_1^* < 0$ corresponding to a positive Newton's coupling. The coefficients $g_n^*, n > 2$ are significantly smaller. E.g., $g_3^*/g_2^* \approx 10^{-3}$ and the numerical values of further coefficients rapidly approaches zero. Thus the solutions $\varphi(r)$ are essentially second order polynomials in the dimensionless curvature $r$.

\begin{figure}[t!]
\centering
\includegraphics[width=0.48\textwidth]{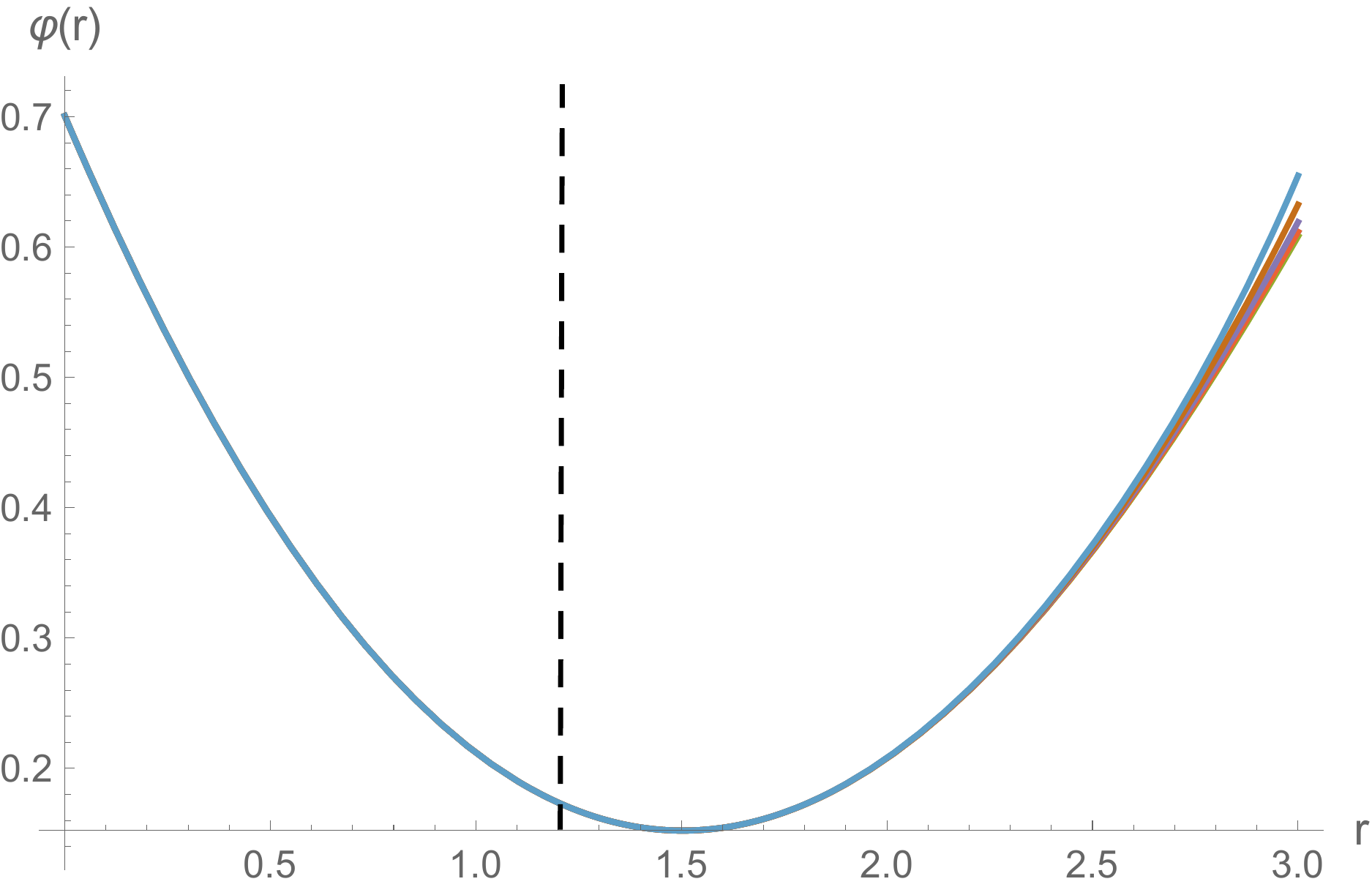} \,
\includegraphics[width=0.48\textwidth]{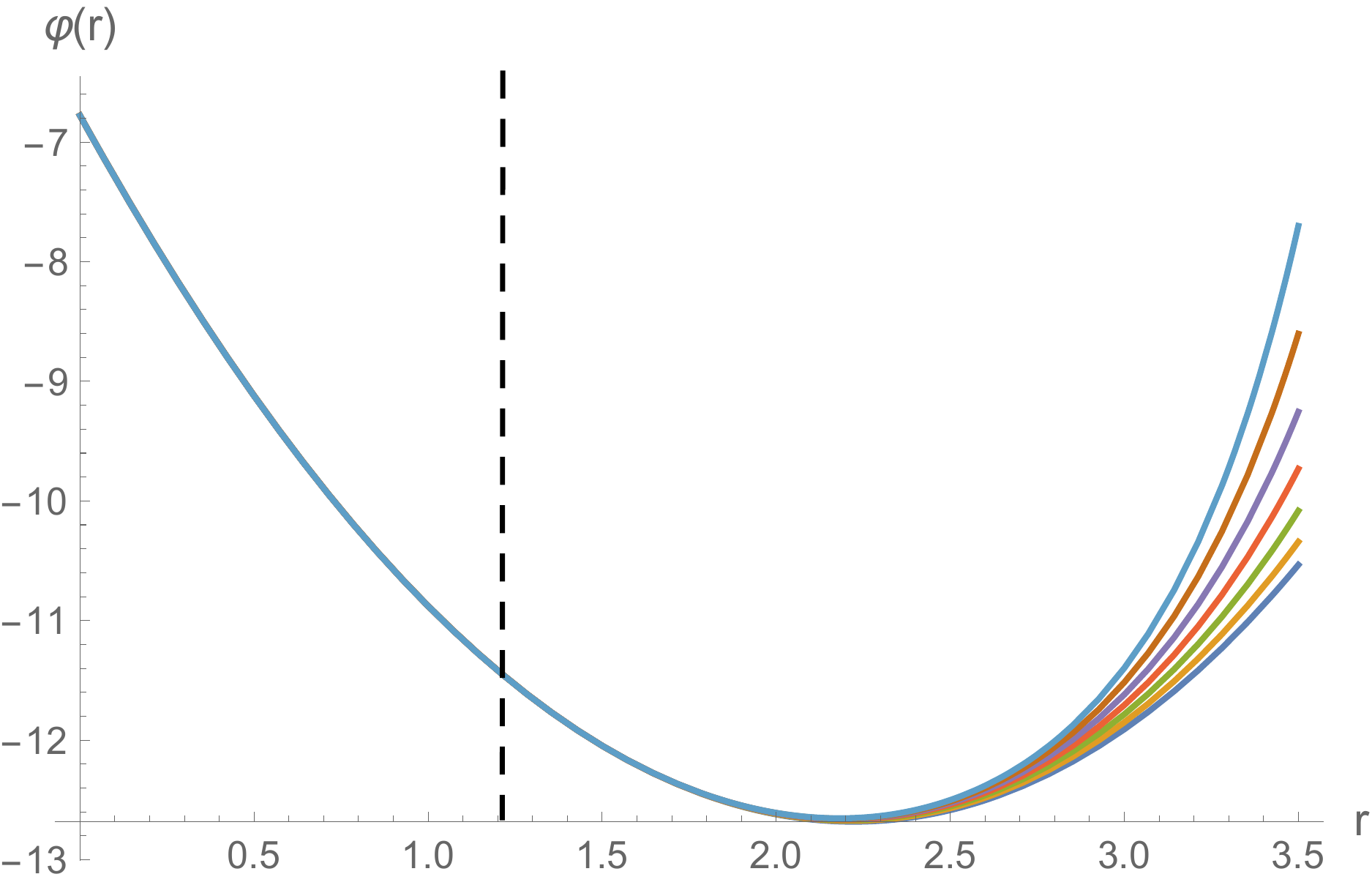}
\caption{\label{Fig.conv} Fixed functions  arising from the polynomial expansion of $\varphi(r)$ for a Type I coarse graining operator. The cases of pure gravity and gravity coupled to the matter content of the standard model are shown in the left and right diagram, respectively. From bottom to top the curves result from the expansions up to $N=8,9,10,11,12,13,14$. The polynomial approximation provides a convergent solution of the fixed point equation which extends up to the moving singularity where $\varphi^\prime(r) = 0$.
}
\end{figure}
The polynomials $\varphi(r)$ for increasing values of $N$ arising for pure gravity and gravity coupled to the matter content of the standard model are shown in the left and right diagram of Fig.\ \ref{Fig.conv}, respectively. For small values of $r$ the polynomial expansion shows a rapid convergence. Notably, both  solutions exhibit a local minimum at $r \approx 1.50$ and $r \approx 2.2$, respectively. Inspecting \eqref{gravTT}, one finds that this minimum corresponds to a moving singularity. In order for the fixed functional to extend to a global solution the zero of $\varphi^\prime$  must be canceled by a corresponding zero in the numerator. For the Type I case where $\alpha_T^G=0$ such a cancellation occurs automatically at $r=3/2$. The interplay between the moving singularity and this cancellation in the case of pure gravity (right diagram of Fig.\ \ref{Fig.conv}) then leads to a polynomial solution whose convergence properties are better than expected on the grounds of the moving singularity.\footnote{Note that the radius of convergence displayed in Fig.\ \ref{Fig.conv} is independent of the fixed singularities given in \eqref{fixedsing}. The construction of the polynomial solution is not based on the normal form of the fixed point equation so that these singular loci are irrelevant for determining the convergence structure of the solution.}

\bigskip \bigskip 

\section{Summary and conclusions}
\label{sec:conclusions}
In this work we reported on a study of the properties of non-Gaussian renormalization group fixed points (NGFPs) arising within $f(R)$-gravity minimally coupled to an arbitrary number of scalar, Dirac, and vector fields. The construction closely follows earlier work by Ohta, Percacci, and Vacca \cite{Ohta:2015efa,Ohta:2015fcu} covering the case of pure gravity: metric fluctuations are parameterized by the exponential split, the computation is carried out in physical gauge, and all operator traces are evaluated as averaged sums over eigenvalues. The result is the partial differential equation \eqref{pdf4d} which governs the scale-dependence of the dimensionless function $\varphi_k(r) \equiv f_k(R/k^2) k^{-4}$. The equation keeps track of a 7-parameter family of coarse graining operators parameterizing relative shifts of the momenta $p^2$ specifying which fluctuations are integrated out at the RG scale $k$. A direct consequence of the construction is that the gravitational sector of our partial differential equation agrees with \cite{Ohta:2015efa,Ohta:2015fcu}. 

Based on the partial differential equation \eqref{pdf4d}, our work develops a comprehensive picture detailing the existence and stability of interacting renormalization group fixed points in gravity-matter systems taking higher-order curvature terms into account.  Our main findings are summarized in Table \ref{Tab.mainresults}. In the case where all coarse graining operators are taken as the corresponding Laplacian operators, most of the matter sectors of phenomenological interest, including the standard model of particle physics, admit a NGFP which is stable under the inclusion of higher-order curvature terms and comes with a low number of relevant directions. The fact that these gravity-matter fixed points share many of the properties found in the case of pure gravity suggests to call this family of universality classes ``gravity-type'' non-Gaussian fixed points. The existence of this class of NGFPs is highly encouraging for working towards a unified picture of all fundamental forces within the framework of asymptotic safety.

In contrast to this success, the most commonly used set of non-trivial endomorphism parameters, given by the Type II coarse graining operators constructed from \eqref{endtypeII}, commonly leads to gravity-matter fixed points which are unstable under the addition of higher order scalar curvature terms. While the instability of phenomenologically interesting gravity-matter fixed points in the presence of a Type II coarse graining operator has already been observed several times,
see, {\it e.g.}, \cite{Dona:2012am,Dona:2013qba}, the present setup offers a striking explanation: the inclusion of the $r^2$-terms reveals that the ``gravity-type'' NGFPs and the NGFPs found in the Type II case \emph{are not connected} by a continuous deformation of the coarse graining operator, see Fig.\ \ref{Fig.typeIIannihilation}. The observation that the matter contributions destroy the typical behavior found in the case of pure gravity suggests to refer to this family of fixed points ``matter-dominated'' NGFPs.

At this stage it is interesting to compare the classification of NGFPs in the Einstein-Hilbert action obtained in this work (see Fig.\ \ref{fig.3}) with the one reported in \cite{Biemans:2017zca}.\footnote{Applying the approach taken in \cite{Biemans:2017zca} to the covariant setting leads to the same existence criteria for NGFPs in the $d_g$-$d_\lambda$-plane \cite{RSunpublished}.} The comparison reveals a qualitative difference in the fixed point structure for $d_g > 0, d_\lambda > 0$. Focusing on the Type I case, our work shows no suitable fixed points beyond the line $d_g = 179/12$ while \cite{Biemans:2017zca} identifies
suitable NGFPs in this region provided that $d_\lambda$ is sufficiently positive. This difference is related to the occurrence of the cosmological constant on the right-hand-side of the flow equation, manifesting itself in terms of denominators of the form $(1-c \lambda)$ with $c$ being a positive number. 
The resulting poles have been linked to a mechanism of gravitational catalysis \cite{Wetterich:2017ixo}.  
The comparison between this work and \cite{Biemans:2017zca} then reveals that these terms also play a crucial role in stabilizing the NGFPs appearing in the upper-right corner of the $d_g$-$d_\lambda$ plane. Notably the NGFPs found in the case of pure gravity or gravity coupled to standard model matter are not located in this region so that the stabilization mechanism is not required to work in these cases.

Our work then poses two natural questions. First, one may want to understand under which conditions the interacting gravity-matter fixed points seen in Sect.\ \ref{sec:fixedpoints} extend to full-fledged fixed functionals constituting global, stationary solutions of \eqref{pdf4d}. Secondly, one may wonder if a change of a coarse-graining operator by a non-trivial endomorphism parameter still corresponds to quantizing the same theory. These points will be addressed in two separate works \cite{Alkofer:inprep} and \cite{us1}, respectively.


\bigskip

\section*{Acknowledgments}
We thank  Holger Gies, Daniel F.\ Litim, 
Jan M.\ Pawlowski and Roberto Percacci for inspiring discussions.
We are grateful to Lando Bosma for cross-checking some of the calculations and Benjamin Knorr for comments on the manuscript.
The work of N.A.\ is supported by the Netherlands Organization for Scientific
Research (NWO) within the Foundation for Fundamental Research on Matter (FOM) grants
13PR3137. 

\bigskip

\begin{appendix}
\section{The middle-of-the-staircase and the Euler-MacLaurin interpolation}
\label{App.A}
The evaluation of the flow equation in Sect.~\ref{sec:spectrum} is based on
averaging over the upper staircase and lower staircase interpolations. In this appendix, we discuss two alternative interpolation schemes. The ``middle-of-the-staircase'' interpolation discussed in Appendix \ref{App.A1} performs the sums at the averaged eigenvalues, setting $p^{(s)} = \tfrac 1 2$. The Euler-MacLaurin interpolation introduced in Appendix \ref{App.A2} replaces the sum by a continuous integral and neglects the discrete correction terms.
\subsection{The middle-of-the-staircase interpolation}
\label{App.A1}
By definition, the middle-of-the-staircase interpolation evaluates the spectral sums \eqref{mattersums} and \eqref{gravitonsum} on the average of the eigenvalues bounding a plateau of the staircase. The resulting values $N^{(s)}_{\rm max}$ are given by \eqref{Ncutoff} evaluated for
\be\label{intfct}
p^{(s)} = \tfrac 1 2 \, , \qquad q^{(s)} = 0 \, . 
\ee
The analogue of (\ref{matterevaluated}) 
and (\ref{gravevaluated}) for this interpolation scheme
is obtained from the replacements
\be\label{reprules}
T_d^{(s)}(N) \to S_d^{(s)}(N^{(s)}_{\rm max}) \, , \qquad
\widetilde{T}_d^{(s)}(N) \to \widetilde{S}_d^{(s)}(N^{(s)}_{\rm max}) \, , 
\ee 
with $N^{(s)}_{\rm max}$ defined in \eqref{Ncutoff} and evaluated at \eqref{intfct}. Notably, the middle-of-the-staircase scheme also removes all non-analytic terms in $r$.

 Comparing the spectral sums resulting from this interpolation scheme to the early-time expansion of the heat-kernel one (again) finds a deviation in the linear term. The two terms can be brought into agreement by setting $q^{(s)} = \tfrac{1}{3} (d-1)$.
 The signs of these parameters are opposite to the corresponding 
 ones obtained for the averaging approximation, (\ref{qopt}). Phrased differently, the linear terms found in the averaging and the  middle-of-the-staircase interpolations differ from
 the corresponding heat-kernel results in opposite directions.  
 
 For completeness we present here the results for the traces (\ref{gravflow}) and (\ref{matterflow}) obtained within the middle-of-the-staircase 
 interpolation. They are given by
 \begin{subequations}\label{gravflowI}
 	\begin{align}
 	\cT^{\rm TT} = & \,
 	\tfrac{5}{2 (4\pi)^2} \, \tfrac{1}{1 + \left(\alpha^G_T + \tfrac{1}{6}\right)r} \left(1 + \left(\alpha^G_T - \tfrac{19}{48}\right)r\right)\left(1 + \left(\alpha^G_T + \tfrac{1}{48}\right)r\right) \\ \nonumber & \; \;
 	+ \tfrac{5}{12 (4\pi)^2} \, \tfrac{\dot{\varphi}^\prime + 2 \varphi^\prime - 2 r \varphi^{\prime\prime}}{\varphi^\prime \left( 1 + \left(\alpha^G_T + \tfrac{1}{6}\right)r \right) } 
 	\left(1 + \left(\alpha^G_T - \tfrac{19}{48}\right)r\right)
 	\left(1 + \left(\alpha^G_T - \tfrac{1}{24}\right)r\right) 
 	\left(1 + \left(\alpha^G_T + \tfrac{1}{48}\right)r\right) \, , \\
 	\cT^{\rm sinv} = & \,
 	\tfrac{1}{2 (4\pi)^2}
 	\tfrac{ \varphi^{\prime\prime}}{\left(1+ \left(\alpha^G_S - \tfrac{1}{3}\right)r\right) \varphi^{\prime\prime} + \tfrac{1}{3} \varphi^\prime}
 	\left(1 + \left(\alpha^G_S - \tfrac{9}{16}\right)r\right)
 	\left(1 + \left(\alpha^G_S + \tfrac{41}{48}\right)r\right)
 	\\ \nonumber & \; \;
 	+ \tfrac{1}{12 (4\pi)^2}
 	\tfrac{\dot{\varphi}^{\prime\prime} - 2 r \varphi^{\prime\prime\prime}}{\left(1+ \left(\alpha^G_S - \tfrac{1}{3}\right)r\right) \varphi^{\prime\prime} + \tfrac{1}{3} \varphi^\prime}
 	\left(1 + \left(\alpha^G_S - \tfrac{9}{16}\right)r\right)  \times \\ 
 	& ~ \hspace{50mm} 
 	\left(1 + \left(2\alpha^G_S + \tfrac {55}{48} \right)r +  
 	\left((\alpha^G_S)^2  + \tfrac {55}{48}  \alpha^G_S - \tfrac{865}{1152} \right) r^2 
 	\right) \, , \nonumber
 	\\
 	\cT^{\rm ghost} = & \, - \tfrac{3}{2 (4\pi)^2} \, \tfrac{1}{1 + (\alpha^G_V - \tfrac{1}{4})r} \, \left( 1 + \left( \alpha^G_V - \tfrac {23}{48} \right) r  \right) 
 	\left( 1 + \left( \alpha^G_V + \tfrac {29}{48} \right) r  \right) \, ,
 	\end{align}
 \end{subequations}
 for the gravity part and
 \begin{subequations}\label{matterflowM}
 	\begin{align}
 	\cT^{\rm scalar} = & \, \frac{N_S}{2 (4\pi)^2} \, \frac{1}{1 + \alpha_S^M r} \,  \left( 1 + \left( \alpha_S^M + \tfrac{5}{48} \right) r \right) \left(1 + \left( \alpha_S^M +  \tfrac{3}{16} \right) r\right) \, , \\ \label{Tdirac1}
 	\cT^{\rm Dirac} = & \, - \frac{2 N_D}{(4\pi)^2} \, \frac 1 {1 + \left( \alpha_D^M + \tfrac 1 4 \right) r}
 	\left(1 + \left(\alpha^M_D + \tfrac{1}{16}\right)r\right) 
 	\left(1 + \left(\alpha^M_D + \tfrac{11}{48}\right)r\right) \, , \\
 	\cT^{\rm vector} = & \, \frac{N_V}{2 (4\pi)^2} \, \bigg(
 	\tfrac{3}{1 + \left( \alpha^M_{V_1} + \tfrac{1}{4} \right) r}
 	\left(1 + \left(\alpha^M_{V_1} - \tfrac{1}{16} \right) r \right)
 	\left(1 + \left(\alpha^M_{V_1} + \tfrac{3}{16} \right)r\right) \\ & \qquad \qquad \nonumber
 	- \tfrac{1}{1 + \alpha^M_{V_2} r} \left( 1 + (\alpha^M_{V_2} + \tfrac{7}{16}) r \right) 
 	\left(1 + ( \alpha^M_{V_2} - \tfrac{7}{48}) r\right)
 	\bigg) \, 
 	\end{align}
 \end{subequations}
 for the matter part. Note that, in contrast to (\ref{Tdirac}), the middle-of-the-staircase interpolation does not lead to a cancellation between the numerator and denominator in the fermion sector. Comparing the expressions obtained from the two interpolation schemes clearly shows that the procedures of summing and averaging do not commute: the trace contributions obtained from summing first and averaging afterwards (averaging interpolation) differ from averaging first and summing afterwards (middle-of-the-staircase interpolation).

\goodbreak
 
\subsection{The Euler-MacLaurin interpolation}
\label{App.A2}
A third interpolation scheme which avoids non-analytic terms in the spectral sums is provided by the Euler-MacLaurin interpolation see, {\it e.g.},  \cite{Dona:2012am}. In this case the finite sums are approximated through the Euler-MacLaurin formula,
\be
\sum_{l=n}^m \, f(l) = \int_n^m dl \, f(l) + \ldots \, , 
\ee
and neglecting the discrete terms. Applying this strategy to the spectral sums \eqref{mattersums} and \eqref{gravitonsum}, identifying $N_{\rm max}^{(s)}$ with \eqref{Ncutoff} based on the values
\be\label{intfct2}
p^{(s)} = 0 \, , \qquad q^{(s)} = 0 \, , 
\ee
leads to replacement rules similar to \eqref{reprules}. By construction, the terms appearing at zeroth and first order in the scalar curvature agree with the early-time expansion of the heat-kernel.

For completeness, we also give the explicit expression for the operator traces entering into \eqref{pdf4d} based on the Euler-MacLaurin interpolation:
\begin{subequations}\label{gravflowM}
        \begin{align}
        \cT^{\rm TT} = & \,
        \tfrac{5}{2 (4\pi)^2} \, \tfrac{1}{1 + \left(\alpha^G_T + \tfrac{1}{6}\right)r} \left(1 + \left(\alpha^G_T - \tfrac{2}{3}\right)r\right)\left(1 + \left(\alpha^G_T + \tfrac{1}{3}\right)r\right) \\ \nonumber & \; \;
                + \tfrac{5}{12 (4\pi)^2} \, \tfrac{\dot{\varphi}^\prime + 2 \varphi^\prime - 2 r \varphi^{\prime\prime}}{\varphi^\prime \left( 1 + \left(\alpha^G_T + \tfrac{1}{6}\right)r \right) } 
\left(1 + \left(\alpha^G_T - \tfrac{2}{3}\right)r\right)^2 
\left(1 + \left(\alpha^G_T + \tfrac{5}{6}\right)r\right) \, , \\
        \cT^{\rm sinv} = & \,
        \tfrac{1}{2 (4\pi)^2}
        \tfrac{ \varphi^{\prime\prime}}{\left(1+ \left(\alpha^G_S - \tfrac{1}{3}\right)r\right) \varphi^{\prime\prime} + \tfrac{1}{3} \varphi^\prime}
        \left(1 + \left(\alpha^G_S + \tfrac{7}{6}\right)r\right)
        \left(1 + \left(\alpha^G_S - \tfrac{5}{6}\right)r\right)
         \\ \nonumber & \; \;
         + \tfrac{1}{12 (4\pi)^2}
         \tfrac{\dot{\varphi}^{\prime\prime} - 2 r \varphi^{\prime\prime\prime}}{\left(1+ \left(\alpha^G_S - \tfrac{1}{3}\right)r\right) \varphi^{\prime\prime} + \tfrac{1}{3} \varphi^\prime}
         \left(1 + \left(\alpha^G_S - \tfrac{5}{6}\right)r\right)^2 \left(1 + \left(\alpha^G_S + \tfrac{13}{6}\right)r\right),  \\ 
        %
        \cT^{\rm ghost} = & \, - \tfrac{3}{2 (4\pi)^2} \, \tfrac{1}{1 + (\alpha^G_V - \tfrac{1}{4})r} \, \left( 1 + \left( \alpha^G_V - \tfrac {3}{4} \right) r  \right) 
\left( 1 + \left( \alpha^G_V + \tfrac {11}{12} \right) r  \right) \, ,
        \end{align}
\end{subequations}
and
\begin{subequations}
        \begin{align}
\cT^{\rm scalar} = & \, \frac{N_S}{2 (4\pi)^2} \,  
\left( 1 + \left( \alpha_S^M + \tfrac{1}{3} \right) r \right)  \, , \\ 
\label{Tdirac2}
\cT^{\rm Dirac} = & \, - \frac{2 N_D}{(4\pi)^2} \, \frac 1 {1 + \left( \alpha_D^M + \tfrac 1 4 \right) r}
\left(1 + \left(\alpha^M_D - \tfrac{1}{12}\right)r\right) 
\left(1 + \left(\alpha^M_D + \tfrac{5}{12}\right)r\right) \, , \\
\cT^{\rm vector} = & \, \frac{N_V}{2 (4\pi)^2} \, \bigg(
\tfrac{3}{1 + \left( \alpha^M_{V_1} + \tfrac{1}{4} \right) r}
\left(1 + \left(\alpha^M_{V_1} - \tfrac{1}{4} \right) r \right)
\left(1 + \left(\alpha^M_{V_1} + \tfrac{5}{12} \right)r\right) \\ & \qquad \qquad \nonumber
- \tfrac{1}{1 + \alpha^M_{V_2} r} \left( 1 + (\alpha^M_{V_2} + \tfrac{2}{3}) r \right) 
\left(1 + ( \alpha^M_{V_2} - \tfrac{1}{3}) r\right)
\bigg) \, .
        \end{align}
\end{subequations}
Note that in this case a cancellation of numerator and denominator takes place for the scalar matter
field.

\newpage  

\section{Fixed point structure of selected gravity-matter systems}
\label{App.B}
In this appendix we collect the fixed point data for the convergent 
NGFP solutions passing the $f(R)$-stability test in Table \ref{Tab.mainresults}. The results for pure gravity are given in Tables \ref{Tab.grav.I} (Type I coarse graining operator) and \ref{Tab.grav.II} (Type II coarse graining operator). In this case the critical exponents with positive real part coincide with the ones obtained in \cite{Ohta:2015efa,Ohta:2015fcu}. The fixed point data for the gravity-matter models featuring matter sectors based on frequently studied models for physics beyond the standard model are displayed in Tables \ref{Tab.FPSMdm} - \ref{Tab.FPMSSM}, respectively. Throughout the presentation, we give results up to $N=8$, and all gravity-matter fixed points show a rapid convergence in the fixed points' position and stability coefficients. Extended computations along the lines of Table \ref{Tab.FPstandardmodel}, covering the critical exponents up to $N=9$ and the polynomial coefficients of the fixed point solution up to $N=14$, confirm this picture.

Following the discussion related to Fig.\ \ref{Fig.typeIIannihilation}, the stable gravity-matter fixed points for a Type II coarse graining scheme can be understood as a deformation of their Type I counterparts. For the matter sectors listed in Table \ref{Tab.3}, these  deformations do not extend to a coarse graining operator of Type II. Hence our lists of stable gravity-matter fixed points comprise results for the Type I coarse graining operator only.    

\medskip
\begin{table}[h!]
	\centering
	\begin{tabular}{c|ccccccccc}
		$N$ & \hspace*{5mm}  $g_0^*$ \hspace*{4mm}  & \hspace*{4mm} $g_1^*$ \hspace*{4mm}  & \hspace*{4mm} $g_2^*$ \hspace*{2mm} & \;  $g_3^* \times 10^{-3}$ & \hspace*{1mm} $g_4^* \times 10^{-4}$ & \hspace*{1mm}  $g_5^* \times 10^{-4}$  & \hspace*{1mm}  $g_6^* \times 10^{-5}$  &  \hspace*{2mm} $g_7^*$ \hspace*{4mm}  & \hspace*{5mm} $g_8^*$ \hspace*{5mm} \\ \hline \hline
		$1$ & $0.46$ & $-1.24$  \\
		$2$ & $0.70$ & $-0.75$ & $0.27$ \\
		$3$ & $0.69$ & $-0.74$ & $0.26$ & $-2.30$   \\
		$4$ & $0.70$ & $-0.75$ & $0.26$ & $-1.27$ & $-6.33$  \\
		$5$ & $0.70$ & $-0.74$ & $0.26$ & $-1.83$ & $-6.38$ & $-1.04$ \\
		$6$ & $0.70$ & $-0.74$ & $0.26$ & $-1.76$ & $-6.87$& $-1.04$ & $-1.99$ \\ 
		$7$ & $0.70$ & $-0.74$ & $0.26$ & $-1.81$ &  $-6.90$ & $-1.13$ & $-2.08$ & $\approx 0$ \\
		$8$ & $0.70$ & $-0.74$ & $0.26$ & $-1.80$ & $-6.93$ & $-1.14$ & $-2.23$ & $\approx 0$ &  $\approx 0$  \\ \hline \hline
		\multicolumn{10}{c}{}\\
		$N$ & $\theta_0$ & $\theta_1$ & $\theta_2$ & $\theta_3$ & $\theta_4$ & $\theta_5$ & $\theta_6$ & $\theta_7$ & $\theta_8$  \\ \hline \hline
		$1$ & $4$ & $2.78$ \\
		$2$ & $4$ & $2.29$ & $-1.50$  \\
		$3$ & $4$ & $2.00$ & $-1.50$ & $-4.01$ \\
		$4$ & $4$ & $2.17$ & $-1.80$ & $-3.99$ &  $-6.23$ 
		\\
		$5$ & $4$ & $2.10$ & $-1.79$ & $-4.37$ & 
		$-6.26$ & $-8.39$
		\\
		$6$ & $4$ & $2.13$ & $-1.87$ & $-4.41$ &
		$-6.62$ & $-8.39$ & $-10.50$ 
		\\ 
		$7$ & $4$ & $2.11$ & $-1.88$ & $-4.50$ & 
		$-6.70$ & $-8.72$ & $-10.50$ & $-12.57$ \\
		$8$ & $4$ & $2.11$ & $-1.90$ & $-4.52$  & $-6.79$ & $-8.80$ & $-10.79$
		& $-12.57$ & $-14.63$ 
		\\ \hline \hline
	\end{tabular}
	\caption{\label{Tab.grav.I} Fixed point structure of $f(R)$-gravity without matter fields $N_S = N_D = N_V = 0$ and a coarse graining operator of Type I. The value of couplings smaller than $10^{-5}$ is indicated by $\approx 0$.}
\end{table}

\begin{table}[p!]
	\centering
	\begin{tabular}{c|ccccccccc}
		$N$ & \hspace*{5mm}  $g_0^*$ \hspace*{4mm}  & \hspace*{4mm} $g_1^*$ \hspace*{4mm}  & \hspace*{4mm} $g_2^*$ \hspace*{2mm} & \;  $g_3^* \times 10^{-2}$ & \hspace*{1mm} $g_4^* \times 10^{-4}$ & \hspace*{1mm}  $g_5^* \times 10^{-4}$  & \hspace*{1mm}  $g_6^* \times 10^{-4}$  &  \hspace*{2mm} $g_7^*$ \hspace*{4mm}  & \hspace*{5mm} $g_8^*$ \hspace*{5mm} \\ \hline \hline
		$1$ & $0.46$ & $-1.78$  \\
		$2$ & $0.67$ & $-1.21$ & $0.51$ \\
		$3$ & $0.65$ & $-1.07$ & $0.53$ & $-4.29$   \\
		$4$ & $0.64$ & $-1.06$ & $0.54$ & $-4.42$ & $8.17$  \\
		$5$ & $0.64$ & $-1.06$ & $0.54$ & $-4.66$ & $6.88$ & $-6.26$ \\
		$6$ & $0.64$ & $-1.06$ & $0.54$ & $-4.63$ & $3.72$& $-6.53$ & $-1.57$ \\ 
		$7$ & $0.64$ & $-1.06$ & $0.54$ & $-4.67$ &  $2.65$ & $-8.04$ & $-1.92$ & $\approx 0$ \\
    	$8$ & $0.64$ & $-1.06$ & $0.53$ & $-4.68$ & $1.45$ & $-8.51$ & $-2.46$ & $\approx 0$ &  $\approx 0$  \\ \hline \hline
		\multicolumn{10}{c}{}\\
		$N$ & $\theta_0$ & $\theta_1$ & $\theta_2$ & $\theta_3$ & $\theta_4$ & $\theta_5$ & $\theta_6$ & $\theta_7$ & $\theta_8$  \\ \hline \hline
		$1$ & $4$ & $2.75$ \\
		$2$ & $4$ & $1.98$ & $-1.22$  \\
		$3$ & $4$ & $1.74$ & $-1.11$ & $-3.96$ \\
		$4$ & $4$ & $1.83$ & $-1.46$ & $-4.02$ &  $-6.68$ 
		\\
		$5$ & $4$ & $1.75$ & $-1.40$ & $-4.42$ & 
		$-6.67$ & $-9.33$
		\\
		$6$ & $4$ & $1.78$ & $-1.46$ & $-4.40$ &
		$-7.11$ & $-9.15$ & $-12.45$ 
		\\ 
		$7$ & $4$ & $1.76$ & $-1.46$ & $-4.46$ & 
		$-7.10$ & $-9.56$ & $-11.52$ & $-16.72$ \\
		$8$ & $4$ & $1.77$ & $-1.48$ & $-4.47$  & $-7.13$ & $-9.54$ & $-11.84$
		& $-13.74$ & $-23.33$ 
		\\ \hline \hline
	\end{tabular}
	\caption{\label{Tab.grav.II} Fixed point structure of $f(R)$-gravity without matter fields $N_S = N_D = N_V = 0$ and a coarse graining operator of Type II. The value of couplings smaller than $10^{-4}$ is indicated by $\approx 0$.}
\end{table}

\begin{table}[p!]
	\centering
	\begin{tabular}{c|ccccccccc}
		$N$ & \hspace*{5mm}  $g_0^*$ \hspace*{4mm}  & \hspace*{4mm} $g_1^*$ \hspace*{4mm}  & \hspace*{4mm} $g_2^*$ \hspace*{2mm} & \;  $g_3^* \times 10^{-3}$ & \hspace*{1mm} $g_4^* \times 10^{-4}$ & \hspace*{1mm}  $g_5^* \times 10^{-5}$  & \hspace*{1mm}  $g_6^* \times 10^{-5}$  &  \hspace*{2mm} $g_7^*$ \hspace*{4mm}  & \hspace*{5mm} $g_8^*$ \hspace*{5mm} \\ \hline \hline
		$1$ & $-7.17$ & $-5.72$  \\
		$2$ & $-6.64$ & $-5.10$ & $1.11$ \\
		$3$ & $-6.64$ & $-5.15$ & $1.13$ & $4.74$   \\
		$4$ & $-6.64$ & $-5.15$ & $1.13$ & $4.61$ & $-2.58$  \\
		$5$ & $-6.64$ & $-5.15$ & $1.12$ & $4.79$ & $-2.39$ & $2.96$ \\
		$6$ & $-6.64$ & $-5.15$ & $1.12$ & $4.87$ & $-1.87$& $3.83$ & $1.26$ \\ 
		$7$ & $-6.64$ & $-5.15$ & $1.13$ & $5.03$ &  $-1.51$ & $6.48$ & $1.79$ & $\approx 0$ \\
		$8$ & $-6.64$ & $-5.15$ & $1.13$ & $5.13$ & $-1.08$ & $7.79$ & $2.77$ & $\approx 0$ &  $\approx 0$  \\ \hline \hline
		\multicolumn{10}{c}{}\\
		$N$ & $\theta_0$ & $\theta_1$ & $\theta_2$ & $\theta_3$ & $\theta_4$ & $\theta_5$ & $\theta_6$ & $\theta_7$ & $\theta_8$  \\ \hline \hline
		$1$ & $4$ & $2.13$ \\
		$2$ & $4$ & $2.36$ & $-1.77$  \\
		$3$ & $4$ & $2.29$ & $-1.83$ & $-6.20$ \\
		$4$ & $4$ & $2.29$ & $-1.92$ & $-6.08$ &  $-9.52$ 
		\\
		$5$ & $4$ & $2.29$ & $-1.92$ & $-6.12$ & 
		$-9.52$ & $-12.17$
		\\
		$6$ & $4$ & $2.29$ & $-1.91$ & $-6.10$ &
		$-9.50$ & $-12.34$ & $-14.50$ 
		\\ 
		$7$ & $4$ & $2.29$ & $-1.90$ & $-6.08$ & 
		$-9.49$ & $-12.27$ & $-14.82$ & $-16.69$ \\
		$8$ & $4$ & $2.29$ & $-1.90$ & $-6.06$  & $-9.46$ & $-12.27$ & $-14.72$
		& $-17.14$ & $-18.81$ 
		\\ \hline \hline
	\end{tabular}
	\caption{\label{Tab.FPSMdm} Fixed point structure of $f(R)$ gravity coupled to the matter content of the standard model supplemented by one additional dark matter scalar where $N_S=5, N_D=\tfrac{45}{2}, N_V = 12$. The value of couplings smaller than $10^{-5}$ is indicated by $\approx 0$.}
\end{table}
\begin{table}[p!]
	\centering
	\begin{tabular}{c|ccccccccc}
		$N$ & \hspace*{5mm}  $g_0^*$ \hspace*{4mm}  & \hspace*{4mm} $g_1^*$ \hspace*{4mm}  & \hspace*{4mm} $g_2^*$ \hspace*{2mm} & \;  $g_3^* \times 10^{-3}$ & \hspace*{1mm} $g_4^* \times 10^{-4}$ & \hspace*{1mm}  $g_5^* \times 10^{-5}$  & \hspace*{1mm}  $g_6^* \times 10^{-5}$  &  \hspace*{1mm} $g_7^* \times 10^{-5}$  & \hspace*{1mm} $g_8^*$ \hspace*{1mm} \\ \hline \hline
		$1$ & $-8.04$ & $-6.08$  \\
		$2$ & $-7.52$ & $-5.46$ & $1.20$ \\
		$3$ & $-7.53$ & $-5.51$ & $1.22$ & $5.25$   \\
		$4$ & $-7.52$ & $-5.51$ & $1.21$ & $5.11$ & $-2.85$  \\
		$5$ & $-7.53$ & $-5.51$ & $1.21$ & $5.31$ & $-2.64$ & $3.45$ \\
		$6$ & $-7.53$ & $-5.51$ & $1.21$ & $5.40$ & $-2.05$& $4.42$ & $1.48$ \\ 
		$7$ & $-7.53$ & $-5.51$ & $1.21$ & $5.58$ &  $-1.66$ & $7.39$ & $2.06$ & $\approx 0$ \\
		$8$ & $-7.53$ & $-5.51$ & $1.21$ & $5.68$ & $-1.18$ & $8.83$ & $3.16$ & $1.12$ &  $\approx 0$  \\ \hline \hline
		\multicolumn{10}{c}{}\\
		$N$ & $\theta_0$ & $\theta_1$ & $\theta_2$ & $\theta_3$ & $\theta_4$ & $\theta_5$ & $\theta_6$ & $\theta_7$ & $\theta_8$  \\ \hline \hline
		$1$ & $4$ & $2.12$ \\
		$2$ & $4$ & $2.33$ & $-1.62$  \\
		$3$ & $4$ & $2.26$ & $-1.67$ & $-5.93$ \\
		$4$ & $4$ & $2.27$ & $-1.75$ & $-5.83$ &  $-9.24$ 
		\\
		$5$ & $4$ & $2.27$ & $-1.75$ & $-5.85$ & 
		$-9.29$ & $-11.91$
		\\
		$6$ & $4$ & $2.27$ & $-1.74$ & $-5.83$ &
		$-9.23$ & $-12.10$ & $-14.26$ 
		\\ 
		$7$ & $4$ & $2.27$ & $-1.73$ & $-5.80$ & 
		$-9.21$ & $-12.02$ & $-14.60$ & $-16.46$ \\
		$8$ & $4$ & $2.27$ & $-1.72$ & $-5.79$  & $-9.18$ & $-12.01$ & $-14.48$
		& $-16.92$ & $-18.59$ 
		\\ \hline \hline
	\end{tabular}
	\caption{\label{Tab.FPSM3nu} Fixed point structure of $f(R)$ gravity coupled to the matter content of the standard model supplemented by three right-handed neutrinos where $N_S=4, N_D=24, N_V = 12$. The value of couplings smaller than $10^{-5}$ is indicated by $\approx 0$.}
\end{table}
\begin{table}[p!]
	\centering
	\begin{tabular}{c|ccccccccc}
		$N$ & \hspace*{5mm}  $g_0^*$ \hspace*{4mm}  & \hspace*{4mm} $g_1^*$ \hspace*{4mm}  & \hspace*{4mm} $g_2^*$ \hspace*{4mm} & \;  $g_3^* \times 10^{-3}$ \; & \; $g_4^* \times 10^{-4}$ \; &   $g_5^* \times 10^{-5}$  &  $g_6^* \times 10^{-5}$  & \hspace*{4mm} $g_7^*$ \hspace*{4mm} & \hspace*{4mm} $g_8^*$ \hspace*{4mm} \\ \hline \hline
		$1$ & $-7.79$ & $-5.87$  \\
		$2$ & $-7.25$ & $-5.24$ & $1.14$ \\
		$3$ & $-7.25$ & $-5.29$ & $1.15$ & $4.65$   \\
		$4$ & $-7.25$ & $-5.29$ & $1.15$ & $4.51$ & $-2.61$  \\
		$5$ & $-7.25$ & $-5.29$ & $1.15$ & $4.68$ & $-2.43$ & $2.76$ \\
		$6$ & $-7.25$ & $-5.29$ & $1.15$ & $4.75$ & $-1.95$& $3.56$ & $1.16$ \\ 
		$7$ & $-7.25$ & $-5.29$ & $1.15$ & $4.90$ &  $-1.60$ & $6.07$ & $1.66$ & $\approx 0$ \\
		$8$ & $-7.25$ & $-5.29$ & $1.15$ & $5.00$ &  $-1.18$ & $7.33$ & $2.58$ & $\approx 0$ & $\approx 0$  \\ \hline \hline
		\multicolumn{10}{c}{}\\
		$N$ & $\theta_0$ & $\theta_1$ & $\theta_2$ & $\theta_3$ & $\theta_4$ & $\theta_5$ & $\theta_6$ & $\theta_7$ & $\theta_8$  \\ \hline \hline
		$1$ & $4$ & $2.13$ \\
		$2$ & $4$ & $2.36$ & $-1.81$  \\
		$3$ & $4$ & $2.29$ & $-1.88$ & $-6.31$ \\
		$4$ & $4$ & $2.29$ & $-1.97$ & $-6.18$ &  $-9.66$ 
		\\
		$5$ & $4$ & $2.29$ & $-1.97$ & $-6.22$ & 
		$-9.66$ & $-12.33$
		\\
		$6$ & $4$ & $2.29$ & $-1.96$ & $-6.21$ &
		$-9.64$ & $-12.49$ & $-14.67$ 
		\\ 
		$7$ & $4$ & $2.29$ & $-1.95$ & $-6.18$ & 
		$-9.63$ & $-12.43$ & $-14.99$ & $-16.86$ \\
		$8$ & $4$ & $2.29$ & $-1.95$ & $-6.16$  & $-9.59$ & $-12.42$ & $-14.88$
		& $-17.31$ & $-18.99$ 
		\\ \hline \hline
	\end{tabular}
	\caption{\label{Tab.FPSMext3} Fixed point structure of $f(R)$ gravity coupled to the matter content of the standard model supplemented by right-handed neutrinos and two additional scalars where $N_S=6, N_D=24, N_V = 12$. The value of couplings smaller than $10^{-5}$ is indicated by $\approx 0$.}
\end{table}
\begin{table}[p!]
	\centering
	\begin{tabular}{c|ccccccccc}
		$N$ & \hspace*{5mm}  $g_0^*$ \hspace*{4mm}  & \hspace*{4mm} $g_1^*$ \hspace*{4mm}  & \hspace*{4mm} $g_2^*$ \hspace*{4mm} & \;  $g_3^* \times 10^{-4}$ \; & \; $g_4^* \times 10^{-5}$ \; & \hspace*{3mm}  $g_5^*$ \hspace*{4mm} & \hspace*{4mm} $g_6^*$ \hspace*{4mm} & \hspace*{4mm} $g_7^*$ \hspace*{4mm} & \hspace*{4mm} $g_8^*$ \hspace*{4mm} \\ \hline \hline
		$1$ & $-5.67$ & $-2.47$  \\
		$2$ & $-4.64$ & $-1.46$ & $0.28$ \\
		$3$ & $-4.65$ & $-1.52$ & $0.29$ & $3.50$   \\
		$4$ & $-4.64$ & $-1.51$ & $0.29$ & $3.24$ & $-3.32$  \\
		$5$ & $-4.64$ & $-1.51$ & $0.29$ & $3.22$ & $-3.35$ & $\approx 0$ \\
		$6$ & $-4.64$ & $-1.51$ & $0.29$ & $3.18$ & $-3.56$& $\approx 0$ & $\approx 0$ \\ 
		$7$ & $-4.64$ & $-1.51$ & $0.29$ & $3.23$ &  $-3.45$ & $\approx 0$ & $\approx 0$ & $\approx 0$ \\
		$8$ & $-4.64$ & $-1.51$ & $0.29$ & $3.27$ &  $-3.30$ & $\approx 0$ & $\approx 0$ & $\approx 0$ & $\approx 0$  \\ \hline \hline
		\multicolumn{10}{c}{}\\
		$N$ & $\theta_0$ & $\theta_1$ & $\theta_2$ & $\theta_3$ & $\theta_4$ & $\theta_5$ & $\theta_6$ & $\theta_7$ & $\theta_8$  \\ \hline \hline
		$1$ & $4$ & $2.33$ \\
		$2$ & $4$ & $5.18$ & $-4.98$  \\
		$3$ & $4$ & $3.21$ & \multicolumn{2}{c}{$-11.03 \pm3.75i$} \\
		$4$ & $4$ & $3.01$ & \multicolumn{2}{c}{$-11.57 \pm6.40i$} &  $-17.60$ 
		\\
		$5$ & $4$ & $2.93$ & \multicolumn{2}{c}{$-12.44 \pm 7.65i$} & 
		$-16.50$ & $-20.79$
		\\
		$6$ & $4$ & $2.90$ & \multicolumn{2}{c}{$-12.70 \pm 8.36i$} &
		$-17.53$ & $-19.66$ & $-23.06$ 
		\\ 
		$7$ & $4$ & $2.88$ & \multicolumn{2}{c}{$-12.89 \pm 8.76i$} & 
		$-18.02$ & $-19.95$ & $-22.41$ & $-25.20$ \\
		$8$ & $4$ & $2.88$ & \multicolumn{2}{c}{$-12.96 \pm 
			8.96i$} & $-18.11$ & \multicolumn{2}{c}{$-21.38 \pm 0.71i$}
		& $-25.05$ & $-27.31$ 
		\\ \hline \hline
	\end{tabular}
	\caption{\label{Tab.FPMSSM} Fixed point structure of $f(R)$ gravity coupled to the matter content of the minimally supersymmetric standard model (MSSM) where $N_S=49, N_D=61/2, N_V = 12$. The value of couplings smaller than $10^{-6}$ is indicated by $\approx 0$.}
\end{table}

\newpage

~

\newpage

\end{appendix}

\end{document}